\documentclass[10pt]{aastex631}

\usepackage{graphicx}
\usepackage{float}
\usepackage{subfigure}
\usepackage{tabularx}
\usepackage{amsmath}
\DeclareUnicodeCharacter {02BC} {\textquotesingle}

\usepackage{CJKutf8}
\newcommand{\YWchinesename}{{\begin{CJK}{UTF8}{gbsn}(王颖翔)\end{CJK}}}
\newcommand{\TLchinesename}{{\begin{CJK}{UTF8}{gbsn}(李坦达)\end{CJK}}}
\newcommand{\SLchinesename}{{\begin{CJK}{UTF8}{gbsn}(毕少兰)\end{CJK}}}
\newcommand{\YLchinesename}{{\begin{CJK}{UTF8}{gbsn}(李亚光)\end{CJK}}}

\makeatletter
\newcommand*\fsize{\f@size pt\relax}
\makeatother
\usepackage{comment}

\newcommand{\kepler}[0]{\emph{Kepler}}

\newcommand{\apogee}[0]{\emph{APOGEE}}
\newcommand{\lamost}[0]{\emph{LAMOST}}

\newcommand{\gaia}[0]{\emph{Gaia}}
\newcommand{\teff}[0]{$T_{\text{eff}}$}

\newcommand{\logg}[0]{$\log g$}
\newcommand{\Dnu}[0]{$\Delta\nu$}

\newcommand{\numax}[0]{\mbox{$\nu_{\rm max}$}} 

\newcommand{\dpi}[1]{$\Delta\Pi$}

\defcitealias{Li_2022}{Li T. et~al.}
\defcitealias{Li_2017}{Li T. et~al.}
\defcitealias{li2022prescription}{Li Y. et~al.}

\newcommand{\mycitealt}[1]{\citetalias{#1}~\citeyear{#1}}

\begin{document}

\title{Asteroseismic Modeling of 1,153 \kepler\ Red Giant Branch Stars: Improved Stellar Parameters with Gravity-Mode Period Spacings and Luminosity Constraints \footnote{Released on May, 1st, 2023}}



\correspondingauthor{Tanda Li}
\email{litanda@bnu.edu.cn}

\affiliation{Institute for Frontiers in Astronomy and Astrophysics, Beijing Normal University, Beijing 102206, China}
\affiliation{Department of Astronomy, Beijing Normal University, Beijing, 100875, Peopleʼs Republic of China}
\affiliation{School of Physics and Astronomy, The University of Birmingham, UK, B15 2TT}
\affiliation{Sydney Institute for Astronomy (SIfA), School of Physics, University of Sydney, NSW 2006, Australia}
\affiliation{co-first author}

\author[0000-0003-4721-1668]{Yingxiang Wang\YWchinesename}
\affiliation{Institute for Frontiers in Astronomy and Astrophysics, Beijing Normal University, Beijing 102206, China}
\affiliation{Department of Astronomy, Beijing Normal University, Beijing, 100875, Peopleʼs Republic of China}
\affiliation{co-first author}

\author[0000-0001-6396-2563]{Tanda Li\TLchinesename}
\affiliation{Institute for Frontiers in Astronomy and Astrophysics, Beijing Normal University, Beijing 102206, China}
\affiliation{Department of Astronomy, Beijing Normal University, Beijing, 100875, Peopleʼs Republic of China}
\affiliation{School of Physics and Astronomy, The University of Birmingham, UK, B15 2TT}
\affiliation{co-first author}

\author[0000-0002-7642-7583]{Shaolan Bi\SLchinesename}
\affiliation{Institute for Frontiers in Astronomy and Astrophysics, Beijing Normal University, Beijing 102206, China}
\affiliation{Department of Astronomy, Beijing Normal University, Beijing, 100875, Peopleʼs Republic of China}

\author[0000-0001-5222-4661]{Timothy R. Bedding}
\affiliation{Sydney Institute for Astronomy (SIfA), School of Physics, University of Sydney, NSW 2006, Australia}

\author[0000-0003-3020-4437]{Yaguang Li\YLchinesename}
\affiliation{Sydney Institute for Astronomy (SIfA), School of Physics, University of Sydney, NSW 2006, Australia}

\begin{abstract}
This paper reports estimated stellar parameters of 1,153 \kepler{} red giant branch stars determined with asteroseismic modeling. We use radial-mode oscillation frequencies, gravity-mode period spacings, \gaia{} luminosities, and spectroscopic data to characterize these stars.
Compared with previous studies, we find that the two additional observed constraints, i.e., the gravity-mode period spacing and luminosity, significantly improve the precision of fundamental stellar parameters. The typical uncertainties are 2.9\% for the mass, 11\% for the age, 1.0\% for the radius, 0.0039 dex for the surface gravity, and 0.5\% for the helium core mass, making this the best-characterized large sample of red-giant stars available to date.
With better characterizations for these red giants, we recalibrate the seismic scaling relations and study the surface term on the red-giant branch.
We confirm that the surface term depends on the surface gravity and effective temperature, but there is no significant correlation with metallicity.

\end{abstract}



\section{Introduction}

NASA's \kepler{} mission \citep{borucki2008kepler} was launched in 2008 and collected high-quality photometry data from its primary field for four years. The mission allowed studying solar-like oscillations on nearly 20,000 red giants. Studies with the \kepler{} data have significantly advanced our understanding of red giants and established asteroseismology as an essential tool for precisely determining fundamental stellar parameters \citep[see reviews by][]{chaplin2013asteroseismology, hekker2017giant, jackiewicz2021solar}.

The grid-based asteroseismic modeling approach has been widely used to estimate the parameters of stars \citep[e.g.,][]{stello2009radius, kallinger2010asteroseismology, basu2011sounding}. Previous research has demonstrated that accurate fundamental parameters, such as mass, radius, surface gravity, and age, can be determined by modeling the oscillation frequencies of stars \citep[e.g.,][]{metcalfe2010precise}. 
However, it is worth noting that some estimates could be highly model-dependent, so tests for systematic bias are crucial. With this in mind, \cite{gai2011depth} examined the model dependence using three different model grids, which inferred that there is almost no model dependence for inferred values of surface gravity and radius, but estimated masses and ages are model-dependent.
Later on, \cite{silva2015ages} compared seven different pipelines and stated that asteroseismology could characterize main-sequence stars with precisions of $\sim$2\%, $\sim$4\%, and $\sim$10\% for radius, mass, and age, respectively.

Evolved stars with different masses are crowded into a narrow red-giant branch on the H-R diagram, posing the challenge of precisely determining their fundamental parameters. Furthermore, unlike main-sequence solar-like oscillators, red-giant oscillations exhibit non-radial modes of a mixed nature, making mode extraction and identification more difficult. 
In our previous study (\mycitealt{Li_2022}, LI22 hereafter), we used the radial model frequencies to determine the masses and ages of 3,642 \kepler{} red giants. We obtained a median precision of 4.5$\%$ for mass and 16$\%$ for age. 
Fully using all oscillations, including mixed dipole modes, allows even better constraints on fundamental parameters for red giants (\citealt{Kallinger_2008}; \citealt{Deheuvels_2012}; \mycitealt{Li_2017}). Previous research demonstrated that mixed modes can constrain stellar masses and ages to precisions of $\sim$5\% and $\sim$10\%, respectively \citep{2016A&A...591A..99P,2018ApJ...855...16Z,2019AJ....157..245H,2021AJ....162..211H,2021MNRAS.505.2336M}. However, extracting and identifying mixed modes is time-consuming and hence difficult to apply to a large sample of stars. 
To effectively use the seismic information in the mixed modes and improve the modeling inferences, a compromise is to use the period spacing of the gravity dipole modes ($\Delta\Pi$) extracted from the mixed modes.
The value of $\Delta\Pi$ is highly sensitive to the properties of the central core \citep{S.Deheuvels_2022, 2013ApJ...766..118M}. Given the core property is the key to understanding evolved stars, $\Delta\Pi$ can hence provide powerful constraints for stars on the red-giant branch \citep[e.g.,][]{Mosser_2011, Vrard_2016}. 
For instance, the distinction between hydrogen-shell burning giant stars and helium-core burning stars can be made by very different period spacings \citep{Bedding_2011}. \cite{Stello_2013} measured the period spacings for 13,000 \kepler{} targets and classified these stars into various groups such as red giant branch, helium-core burning clump, and secondary clump. 

In this work, we aim to improve our previous seismic determinations for \kepler{} red giants by using the gravity-mode period spacing and \gaia{} luminosity as additional constraints. The rest of the paper is organized as follows: Section \ref{sec:2} describes our data set and modeling approach; Section \ref{sec:3} presents the results; we close with a summary in Section \ref{sec:4}.

\section{Target Selection and Modeling Approach} \label{sec:2}

\subsection{Data} \label{sec:Data}

In this study, the sample of \kepler{} red-giant-branch (RGB) stars in LI22 serves as the basis for our analysis. This initial sample is composed of 3,642 RGB stars with measured radial oscillation mode frequencies. 
It is worth noting that although they are classified as RGBs by \cite{2018MNRAS.476.3233H}, we cannot be certain they are all RGBs, as asymptotic-giant-branch stars would have similar period spacings and be classified as RGB.
We cross-matched the sample with the \kepler{} Red Giant Period Interval Catalog given by \citet[V16, hereafter]{Vrard_2016} to determine period spacings.
Further, we removed stars below the so-called RGB sequence on the \Dnu --$\Delta\Pi$ diagram classified by \citet{S.Deheuvels_2022} and \citet{2021MNRAS.508.1618R}, as these stars are assumed to have undergone mass transfer or merging events.

Utilising the \gaia{} DR3 catalog \citep{gaia-2016-mission,gaia-2020-edr3}, we calculated luminosities, $L$. Given that Gaia parallaxes are known to contain zero-point offsets, we used a model from \citet{2021A&A...649A...4L} to adjust for this offset.
Additionally, the reported parallaxes have underestimated uncertainties. According to external calibrations \citep{2021A&A...649A..13M,2021MNRAS.506.2269E,2021AJ....161..214Z}, we, therefore, increased them by a factor of 1.3.
To correct for extinction, we used the ``direct'' approach in the program {\small ISOCLASSIFY} \citep{2020AJ....159..280B,2017ApJ...844..102H}, which incorporates the \cite{2019ApJ...887...93G} dust map and the bolometric corrections from MIST models \citep{2016ApJ...823..102C}, to determine the luminosities by combining the parallaxes with the 2MASS $K$-band magnitudes.

\begin{figure}[ht!]
    \centering
    \includegraphics{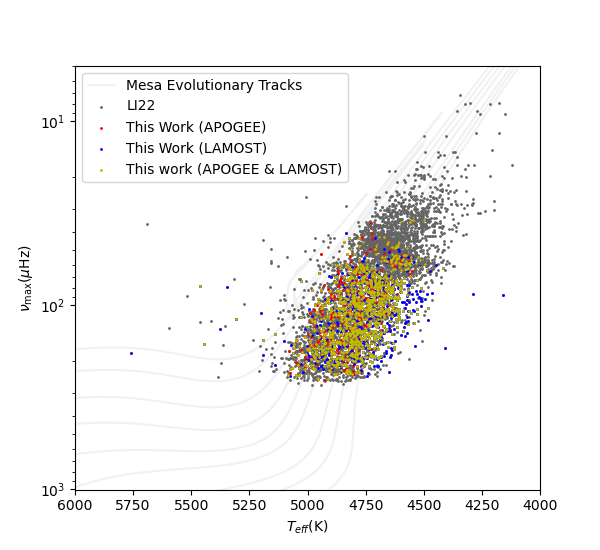}
    \caption{The RGB star samples on the \teff–$\nu_{\rm max}$ diagram. Grey dots indicate the original star sample studied by LI22. The blue and red dots represent the 887 \lamost{} targets and 776 \apogee{} targets studied in this work, of which 510 stars are in both samples.}
    \label{fig:2}
\end{figure}

The final sample includes 1,153 \kepler{} RGB stars, 887 of which are \lamost{}{} targets, 776 are \apogee{}{} targets, and 510 are common sources.
In Figure \ref{fig:2}, we show the sample on the \teff–$\nu_{\rm max}$ diagram. Compared with the original sample, most stars below \numax{} $\sim$ 35 $\mu$Hz are not included in this work because their $\Delta\Pi$ values were not measured.

\subsection{Modeling Approach} \label{subsec: modeling}

We used the stellar model grid calculated by LI22. The grid covers a mass range of 0.76–-2.20 $M_\odot$ with four independent model input parameters: mass ($M$),  initial helium fraction ($Y_{\rm init}$), initial metal abundance ({\rm [M/H]}), and mixing length parameters ($\alpha_{\rm MLT}$). However, our preliminary research demonstrated that the grid's mass resolution (0.02 $M_{\odot}$) is insufficient when we have the two additional observed constraints. For this reason, we computed more models and decreased the mass step to 0.01 $M_\odot$. 

Before further analysis, we tested whether improving the grid resolution would significantly impact the parameter precision given by LI22. We redid the fits of LI22 using the new grid and found only slight decreases in the median uncertainties of mass (from 4.5\% to 4.3\%), radius (from 1.7\% to 1.5\%), and surface gravity (from 0.0062 to 0.0055 dex) and a slightly larger improvement in the age uncertainty (from 16\% to 13\%). This indicates the original grid in LI22 was slightly under-sampled for estimating stellar ages.  

In this work, we adopted the same fitting method described by LI22. We included effective temperature ($T_{\rm eff}$), metallicity ({\rm [M/H]}), and luminosity ($L$) as non-seismic observed constraints and computed the non-seismic likelihood using the following equation:
\begin{equation}
    p_{\rm non-seismic} = {\rm exp}\left[ -\frac{(T_{\rm eff,obs}-T_{\rm eff,mod})^{2}}{{2}(\sigma_{T_{\rm eff,obs}}^{2}+\sigma_{T_{\rm eff,sys}}^{2})}-\frac{({{\rm [M/H]}}_{\rm obs}-{{\rm [M/H]}}_{\rm mod})^{2}}{{2}\sigma_{{{\rm [M/H]}}_{\rm obs}}^{2}}-\frac{(L_{\rm obs}-L_{\rm mod})^{2}}{{2}\sigma_{L_{\rm obs}}^{2}}\right],
\end{equation}
where the subscripts `mod' and `obs' represent the model and observations, respectively. As the reported APOGEE and LAMOST uncertainties are random uncertainties, we add a systematic \teff{} uncertainty (by adopting typical values 2.4\% reported by \cite{2022ApJ...927...31T}) to the random uncertainty in quadrature.
Note that when the observed uncertainty of {\rm [M/H]} was less than 0.1 dex, we used $\sigma_{\rm [M/H_{obs}]}$ = 0.1 dex because of the {\rm [M/H]} grid resolution.
The seismic constraints were the radial mode frequencies and the asymptotic g-mode period spacing ($\Delta\Pi$). The measurement of $\Delta\Pi$ is done using the method by V16, which does not calculate $\Delta\Pi$ after extracting all mixed-mode vibration frequencies. That is, $\Delta\Pi$ does not represent vibration frequencies, so we treat it as an observational measurement,
and we calculated the seismic likelihood function as 
\begin{equation}
    p_{\rm seismic} = {\rm exp}\left[ -\frac{({\Delta}\Pi_{\rm obs}-{\Delta}\Pi_{\rm mod})^{2}}{{2}\sigma_{{\Delta}\Pi_{\rm obs}}^{2}}\right] \times \prod  {\rm exp}\left[ -\frac{(\nu_{i,{\rm obs}}-\nu_{i,{\rm mod}})^{2}}{{2}\sigma_{\nu_{i,{\rm obs}}}^{2}}\right],
\end{equation}
where the subscript $i$ denotes the $i$th mode frequency. 
The final likelihood is $p_{\rm non-seismic} \cdot p_{\rm seismic}$. We estimated each stellar parameter and its uncertainty by measuring the cumulative values at 16$\%$, 50$\%$, and 84$\%$ on the marginal likelihood distribution. 
Note that we adopted the two-term formula and the method proposed by \cite{2014A&A...568A.123B} to correct the surface term in theoretical oscillation frequencies. Corrections to mode frequencies are defined as
\begin{equation}
\label{eq.7}
    \delta\nu=\left(a_{-1}\left(v / v_{\mathrm{ac}}\right)^{-1}+a_{3}\left(v / v_{\mathrm{ac}}\right)^{3}\right) / \mathcal{I}.
\end{equation}
Here, $\nu_{\rm ac}$ is the acoustic cutoff frequency, which is considered to be a fixed fraction of \numax{} \citep{1991ApJ...368..599B, 1995A&A...293...87K}, and $a_{-1}$ and $a_3$ are free parameters. We use the fractional frequency correction at \numax{}, i.e., $\delta\nu(\nu_{\rm max})/\nu_{\rm max}$, to quantity the surface term in stellar models. For each star, we used the same method to estimate $\delta\nu(\nu_{\rm max})/\nu_{\rm max}$ as we used to estimate the stellar parameters.

\section{Results} \label{sec:3}

\subsection{Improved stellar parameters}

\begin{figure}[t]
\centering    
\includegraphics[width=8.5cm]{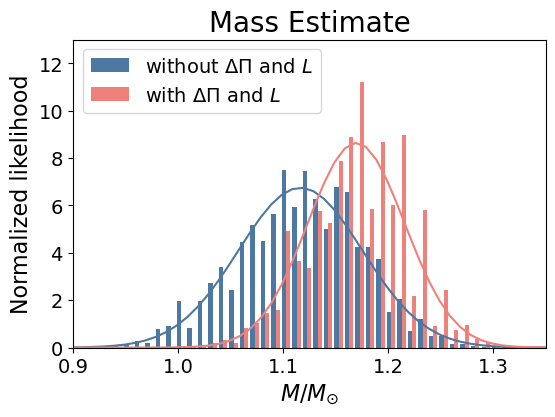}
\includegraphics[width=8.5cm]{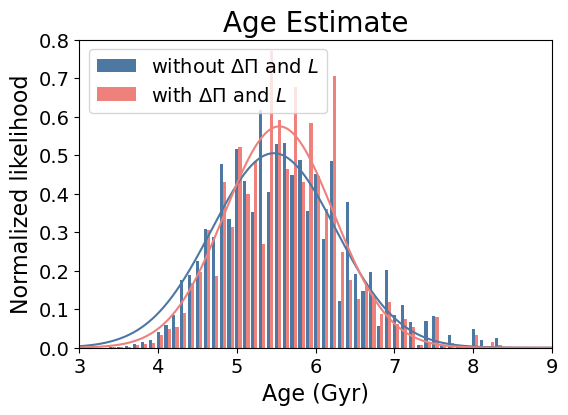}
\includegraphics[width=8.5cm]{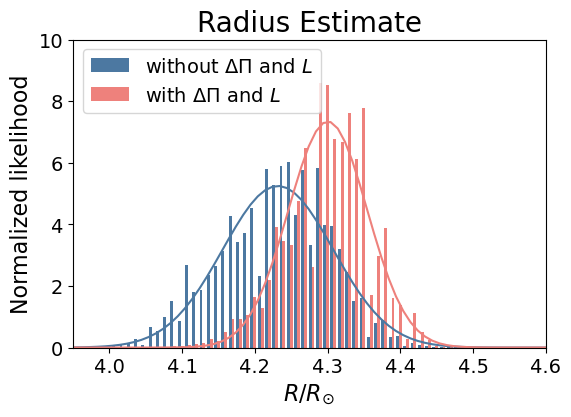}
\includegraphics[width=8.5cm]{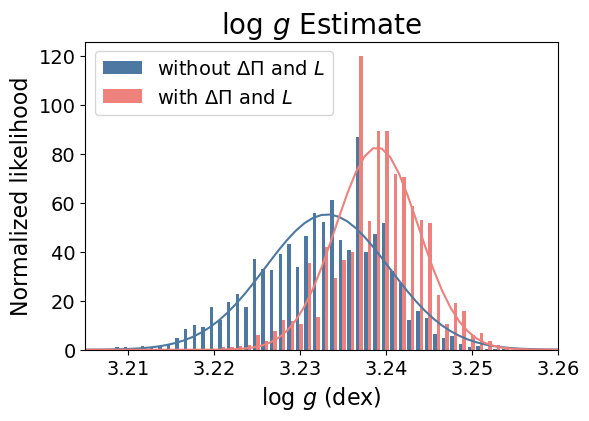}
\caption{Likelihood distributions of mass, age, radius, and surface gravity for KIC 2578581. Blue bars indicate likelihood distributions determined from effective temperature, metallicity, and radial mode frequencies. Red bars represent the likelihood distributions determined with the addition of radial mode frequencies and the g-mode period spacing. The solid curves show Gaussian fits. The metallicity and effective temperature are from the \apogee{} survey.}
\label{fig:1}
\end{figure}  

\begin{table}[b]
\caption{Observed Constraints and Model-inferred Stellar Parameters (The full table is available online.)}
\label{table:OC&MSP}
\centering
\footnotesize
\begin{tabular}{cccccccccccccc}
\hline\hline
        & \multicolumn{7}{c}{Observed Constraints}                                                   &  & \multicolumn{5}{c}{Estimates}                                                                     \\ \cline{2-8} \cline{10-14} 
KIC     & Source & $T_{eff}$ & {[}M/H{]} & $\Delta\nu$ & $\nu_{\rm max}$ & $\Delta\Pi$ & L             &  & M                      & $\tau$              & R                      & log g                     & $m_c$                        \\
        &        & (K)       & (dex)     & ($\mu$Hz)   & ($\mu$Hz)   & (s)           & ($L_{\odot}$) &  & ($M_{\odot}$)          & (Gyr)               & ($R_{\odot}$)          & (dex)                     & ($M_{\odot}$)                \\ \hline
2578581$^{*}$ & APOGEE & 4957      & -0.181     & 15.95        & 209.2        & 85.9          & 11.5          &  & $1.17^{+0.04}_{-0.04}$ & $5.6^{+0.7}_{-0.7}$ & $4.30^{+0.06}_{-0.05}$ & $3.239^{+0.006}_{-0.004}$ & $0.1913^{+0.0014}_{-0.0013}$ \\ 
1027337 & APOGEE & 4636      & 0.231     & 6.94        & 74.2        & 70.1          & 27.3          &  & $1.33^{+0.04}_{-0.03}$ & $5.4^{+0.7}_{-0.7}$ & $7.72^{+0.08}_{-0.06}$ & $2.784^{+0.004}_{-0.003}$ & $0.2247^{+0.0008}_{-0.0008}$ \\
1433803 & APOGEE & 4736      & 0.236     & 12.18       & 150.1       & 79.3          & 12.1          &  & $1.31^{+0.06}_{-0.03}$ & $6.1^{+0.7}_{-1.0}$ & $5.32^{+0.07}_{-0.06}$ & $3.101^{+0.004}_{-0.004}$ & $0.2005^{+0.0011}_{-0.0011}$ \\
1569842 & APOGEE & 4820      & -0.276    & 11.77       & 135.0       & 80.6          & 12.2          &  & $1.03^{+0.02}_{-0.02}$ & $9.0^{+0.6}_{-1.4}$ & $5.01^{+0.04}_{-0.04}$ & $3.051^{+0.003}_{-0.003}$ & $0.1978^{+0.0008}_{-0.0011}$  \\
1723752 & APOGEE & 5011      & -0.155    & 15.04       & 197.3       & 83.7          & 12.6          &  & $1.27^{+0.04}_{-0.04}$ & $4.1^{+0.4}_{-0.5}$ & $4.61^{+0.06}_{-0.05}$ & $3.214^{+0.004}_{-0.004}$ & $0.1963^{+0.0012}_{-0.0010}$ \\
1723843 & APOGEE & 4960      & -0.239    & 9.41        & 108.0       & 72.7          & 26.2          &  & $1.43^{+0.05}_{-0.05}$ & $2.4^{+0.2}_{-0.3}$ & $6.52^{+0.08}_{-0.07}$ & $2.964^{+0.004}_{-0.005}$ & $0.2219^{+0.0013}_{-0.0009}$ \\ \hline
1027337 & LAMOST & 4614      & 0.167     & 6.94        & 74.2        & 70.1          & 27.0          &  & $1.35^{+0.02}_{-0.04}$ & $4.8^{+0.5}_{-1.1}$ & $7.78^{+0.06}_{-0.07}$ & $2.787^{+0.002}_{-0.004}$ & $0.2247^{+0.0009}_{-0.0007}$ \\
1429505 & LAMOST & 4654      & -0.152    & 5.76        & 55.8        & 67.9          & 29.2          &  & $1.12^{+0.03}_{-0.03}$ & $6.8^{+0.5}_{-0.7}$ & $8.24^{+0.08}_{-0.08}$ & $2.654^{+0.005}_{-0.004}$ & $0.2277^{+0.0014}_{-0.0013}$ \\
1433803 & LAMOST & 4690      & 0.176     & 12.18       & 150.1       & 79.3          & 11.8          &  & $1.31^{+0.02}_{-0.02}$ & $5.6^{+0.3}_{-0.3}$ & $5.32^{+0.02}_{-0.03}$ & $3.101^{+0.002}_{-0.004}$ & $0.2003^{+0.0005}_{-0.0007}$ \\
1576646 & LAMOST & 4778      & 0.001     & 7.67        & 84.5        & 69.6          & 24.8          &  & $1.43^{+0.03}_{-0.03}$ & $3.6^{+0.5}_{-0.6}$ & $7.42^{+0.05}_{-0.06}$ & $2.851^{+0.003}_{-0.004}$ & $0.2256^{+0.0008}_{-0.0008}$ \\
1723843 & LAMOST & 4902      & -0.243    & 9.41        & 108.0       & 72.7          & 25.4          &  & $1.47^{+0.03}_{-0.05}$ & $2.5^{+0.2}_{-0.2}$ & $6.59^{+0.05}_{-0.08}$ & $2.968^{+0.003}_{-0.004}$ & $0.2209^{+0.0011}_{-0.0019}$ \\ \hline
\multicolumn{13}{l}{$^*$ The example star in Figure~\ref{fig:2}.}
\end{tabular}
\end{table}

We started by inspecting the improvements in likelihood distributions of stellar parameters due to the two additional constraints. Figure \ref{fig:1} shows comparisons between the likelihood distributions with and without $\Delta\Pi$ and $L$ of an example star (KIC 2578581). The two additional constraints lead to more precise estimates of the four fundamental parameters. In LI22, the precision for the example star was 5.4$\%$ for the mass, 16$\%$ for the age, 0.015 dex for the surface gravity, and 1.8$\%$ for the radius. In this work, the precision is now improved to 3.4$\%$ for mass, 12$\%$ for age, 0.009 dex for surface gravity, and 1.2$\%$ for radius.

\begin{figure}[ht]
 \centering
\includegraphics[width=5.5cm]{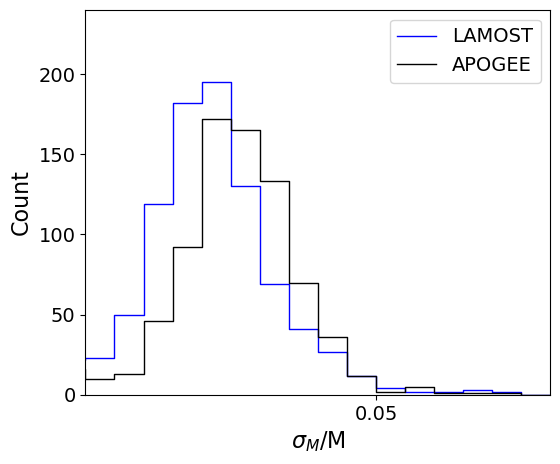}
\includegraphics[width=5.5cm]{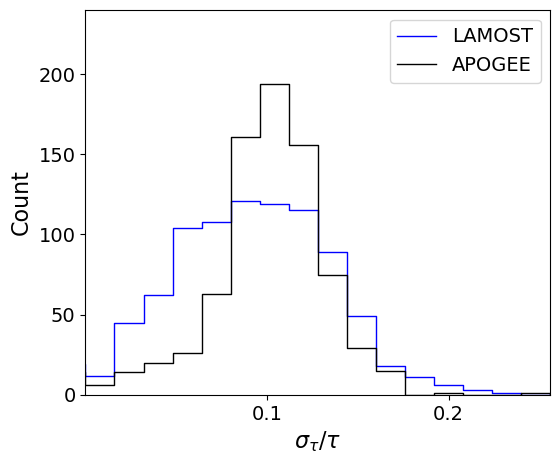}
\includegraphics[width=5.5cm]{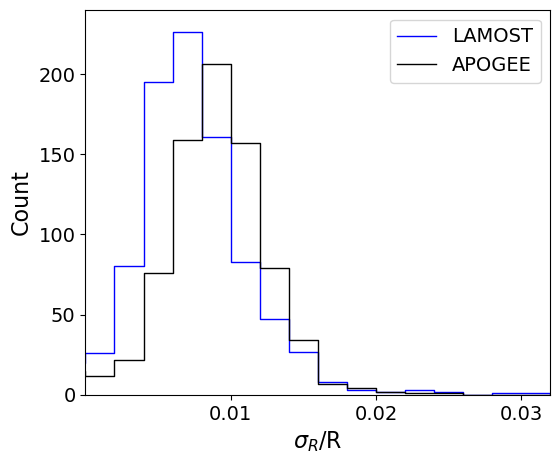}
\includegraphics[width=5.5cm]{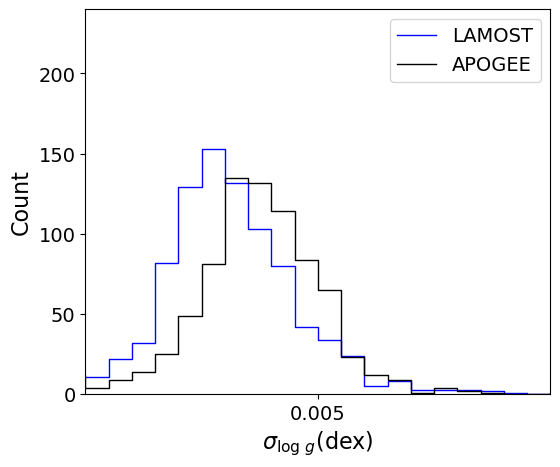}
\includegraphics[width=5.5cm]{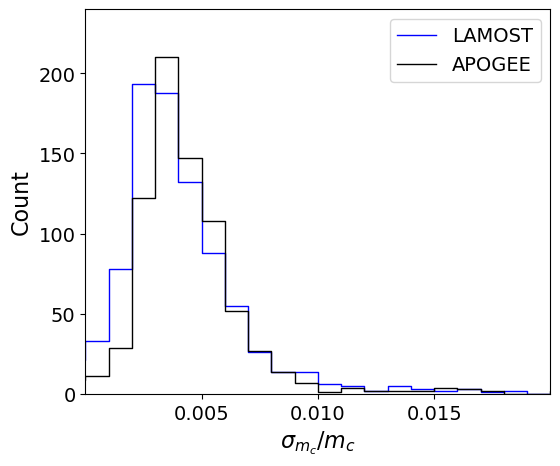}
\caption{Distributions of fractional uncertainties for mass, age, radius, surface gravity, and helium core mass. Blue and black bars indicate star samples with \apogee{} and \lamost{} spectroscopic constraints.}
\label{fig:3}
\end{figure}

We estimated stellar fundamental parameters in this way for the 1,153 RGB stars in our sample. Table~\ref{table:OC&MSP} lists the observed constraints and estimated parameters. (We also provide the information of example star in Figure~\ref{fig:2} at the beginning.) Note there are two entries for the 510 stars that have both \lamost{} and \apogee{} measurements. As found by LI22, there is good agreement between estimated parameters for stars with both \lamost{} and \apogee{} spectroscopic constraints.
Figure~\ref{fig:3} shows the uncertainty distribution of the five stellar parameters. Interestingly, we obtained relatively high precision for the \lamost{} targets because their uncertainties in \teff{} and [M/H] are mostly smaller than those from the \apogee{} survey. Given that the \lamost{} spectra are lower-resolution, the uncertainties may be over-optimistic compared with the \apogee{} high-resolution results and the typical precision determined with the \apogee{} data may be more representative for red giants.   
The median uncertainty of our estimates is 2.9$\%$ for mass, 11$\%$ for age, 1.0$\%$ for radius, and 0.0039 dex for log g. In comparison to the previous results, adding the two new constraints improved the precision by a significant amount. Moreover, we estimated helium core masses, which are also listed in Table~\ref{table:OC&MSP}. The median uncertainty of helium core mass is 0.5$\%$. 

\begin{figure}[ht]
\centering    
\includegraphics[width=8.8cm]{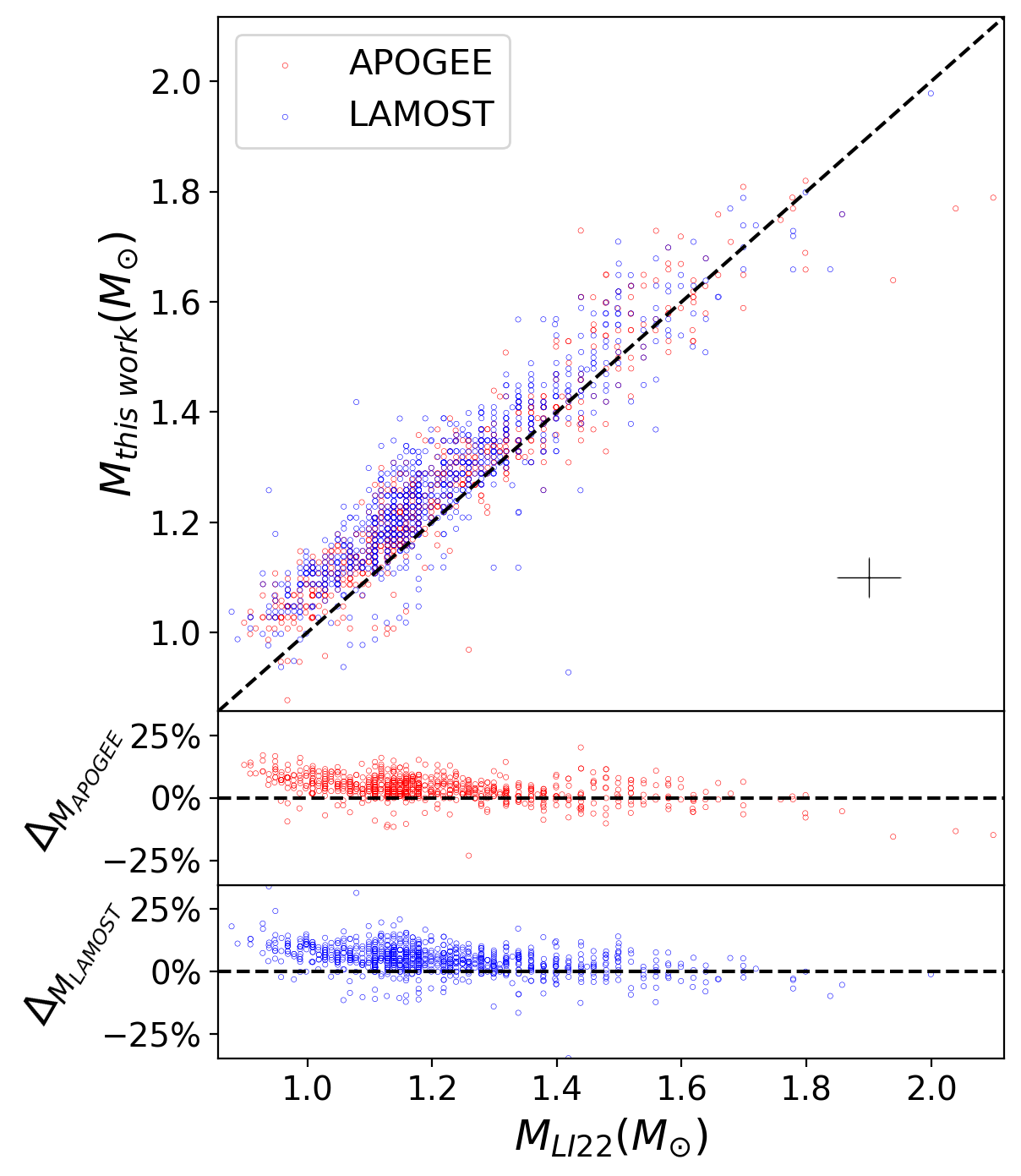}
\includegraphics[width=8.8cm]{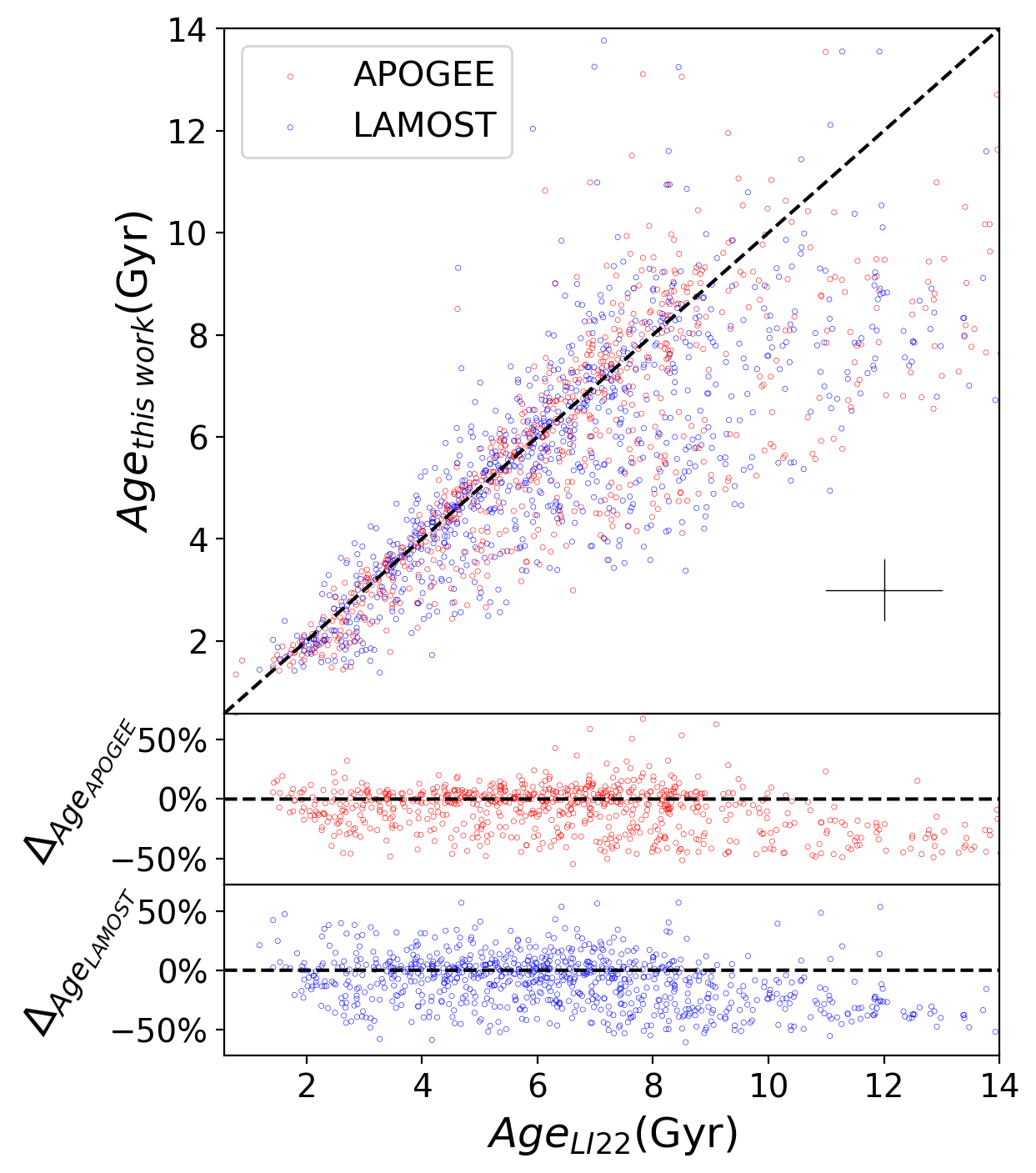}
\includegraphics[width=8.8cm]{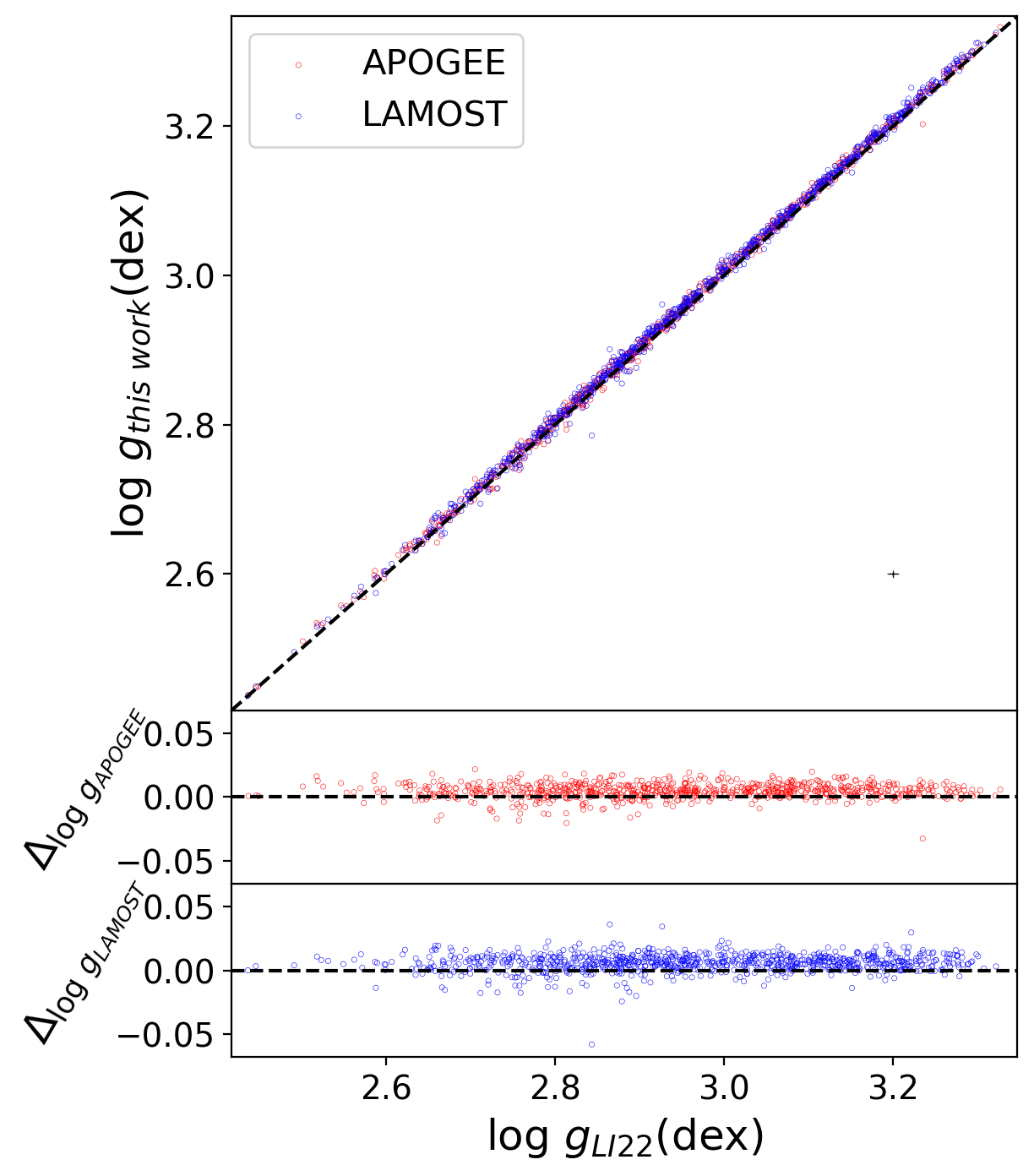}
\includegraphics[width=8.8cm]{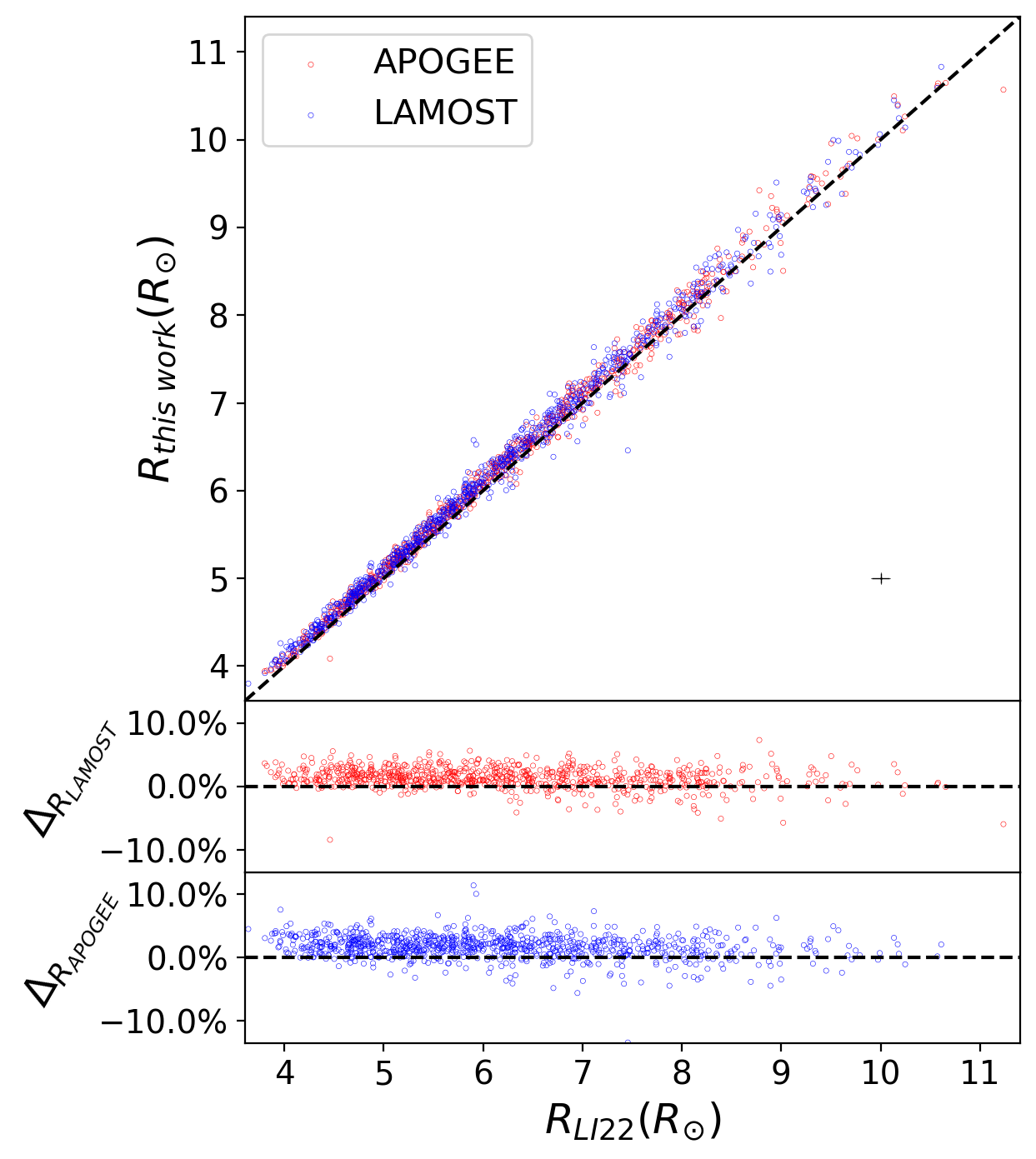}
\caption{Comparison of the fitting results of the fundamental parameters from LI22 and this work. For each subplot (corresponding to a fundamental parameter), the upper part is the comparison of the fitted results, and the middle and the lower parts are the differences in the fitting results for the APOGEE and LAMOST targets, respectively. The blue dots are the targets of LAMOST, and the red dots are the targets of APOGEE. In addition, each figure has a black dot with error bars, the size of which characterizes the average error of the sample.}
\label{fig:4}
\end{figure}

\begin{figure}
    \centering
    \includegraphics[width=8.8cm]{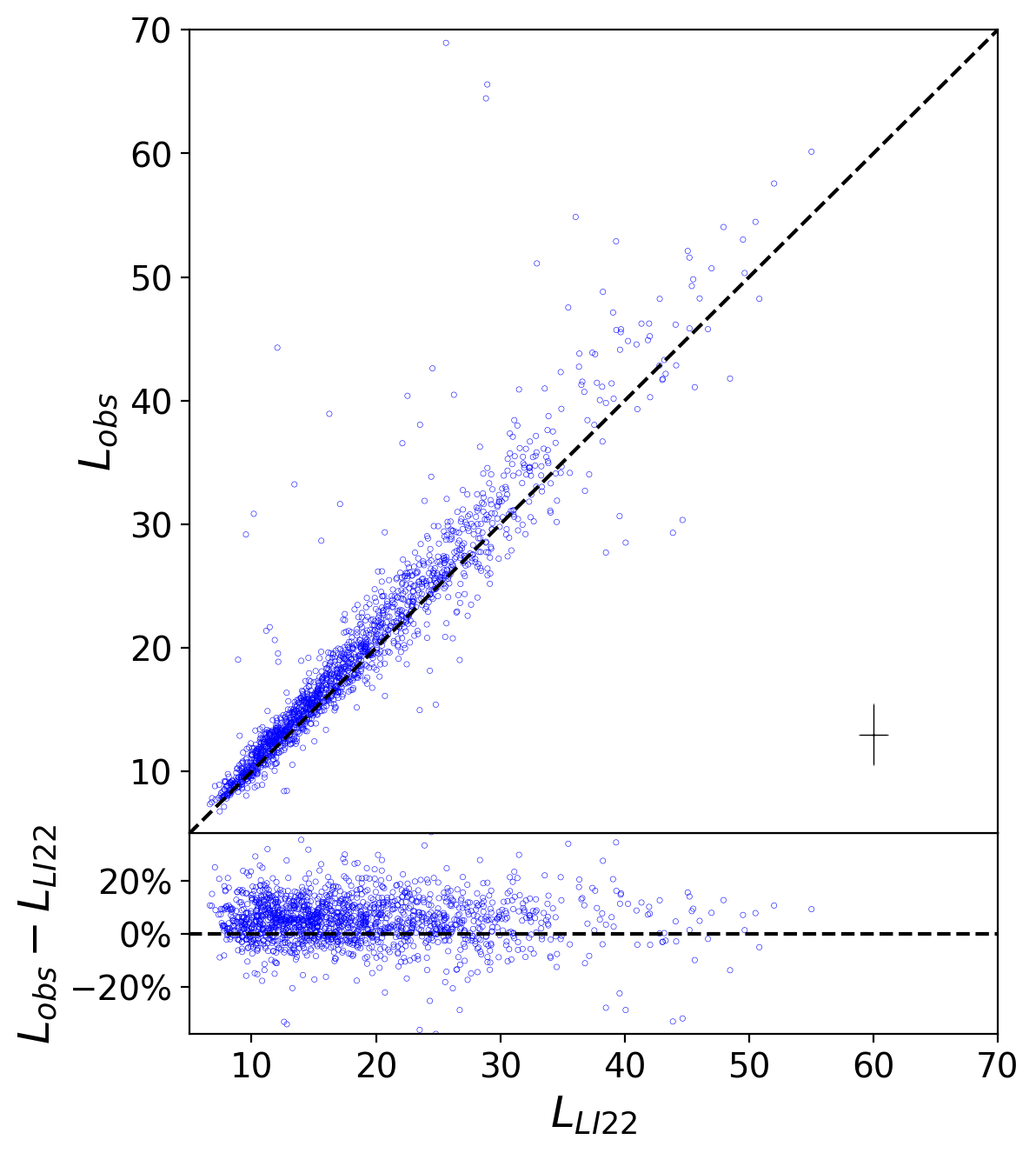}
    \includegraphics[width=8.8cm]{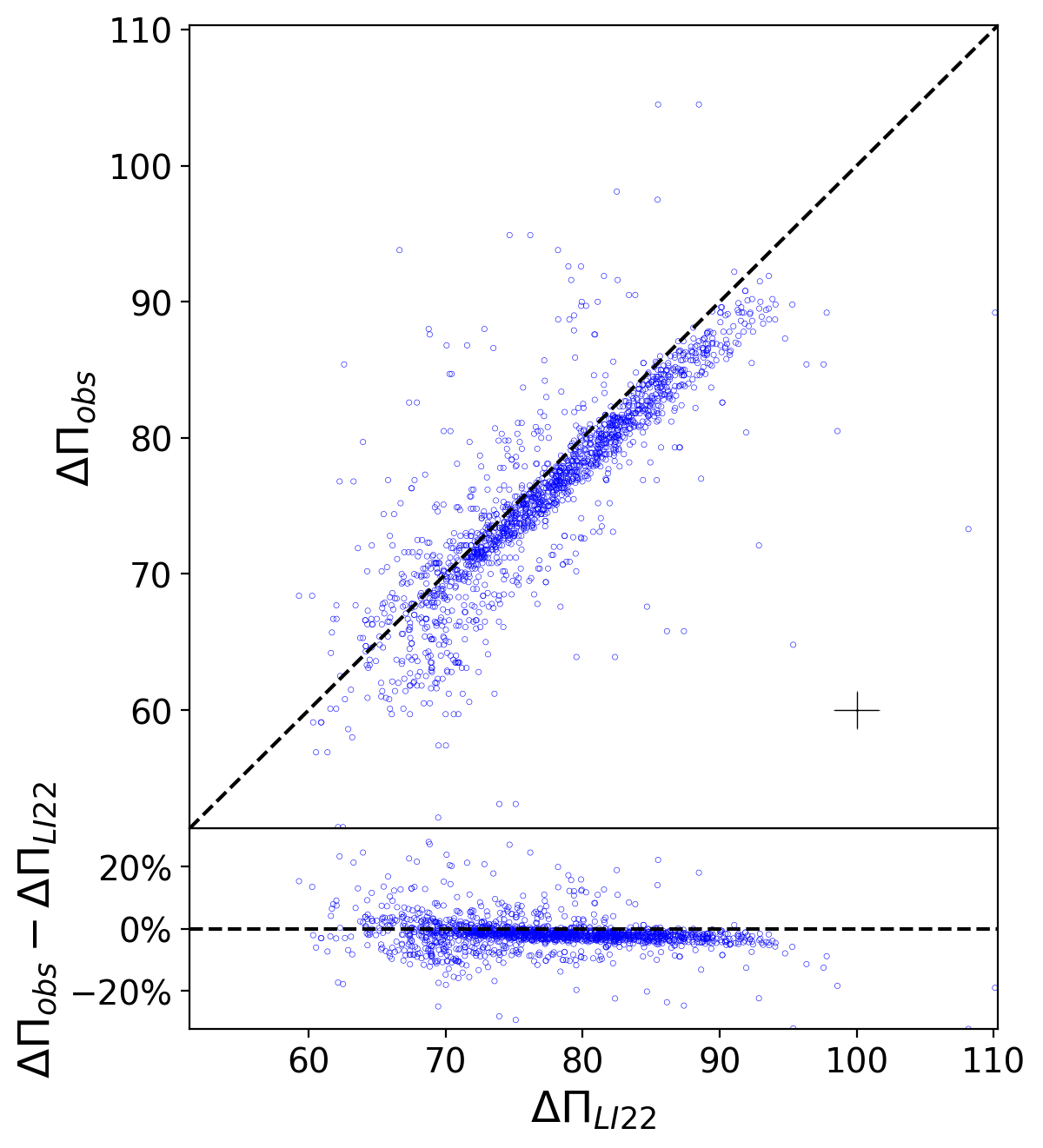}
    \caption{Comparison of observed and fitted values for luminosity and period spacing. The figure shows two scatter plots with error bars, one for each variable, where the x-axis is the result of fitting a stellar oscillation model (LI22) using radial mode frequency and spectral observation quantity, and the y-axis is the observed value from Gaia DR3 and V16 data. This figure reveals the systematic differences between models and observations, which are also the source of the systematic differences in Figure~\ref{fig:4}.}
    \label{fig:5}
\end{figure}

We compared our new determinations with the results in LI22 to examine systematic offsets in Figure~\ref{fig:4}.
We find that our new results suggest slightly higher masses, where the average offset is 4.4\%. The other offsets are $-9.3$\% for age, 0.005 dex for surface gravity, and 1.5\% for radius.
The offsets indicate that the two additional constraints bring in systematic effects. In Figure~\ref{fig:5}, we compared the modeling-inferred luminosities and g-mode period spacings given by LI22 with the observations. Apparent offsets are seen in both parameters. Observed values are slightly large for the luminosity and relatively small for the g-mode period spacing. For red-giant stars, the radius and the surface gravity increase with the luminosity and are inversely proportional to the g-mode period spacing. This explains the relatively large estimates for the surface gravity and the radius. 
It follows that our estimated masses systematically increase due to the increased radii given the fact that the mean densities of these stars are very well constrained by radial mode frequencies.  

This work has noticeably improved parameter precision compared to that of LI22. We also examined the contributions from different aspects. 
Table~\ref{table:uncertainty} lists the median precision with the old and new grids and changes for different observed constraints. 
We note that having luminosity as an additional constraint slightly improves the parameter precision and that the g-mode period spacing (\dpi{}) has a much greater impact on the parameter precision than the luminosity ($L$). Thus, the improvements in parameter precision are mainly from the g-mode period spacing.

\begin{table}[b]
\caption{Changes in median uncertainty estimated by different methods}
\label{table:uncertainty}
\centering
\setlength{\tabcolsep}{25pt}
\begin{tabular}{ccccc}
\hline\hline
        & \multicolumn{4}{c}{Median Precision}                                                                   \\ \cline{2-5}
                                          & M                              & $\tau$                         & log g                              & R                          \\ \hline
LI22 (old grid)                                      & 4.5\% & 16.0\% & 0.0062 & 1.7\% \\
new grid                        & 4.3\%                         & 12.2\%                         & 0.0057                         & 1.5\%                         \\
new grid + \dpi{} as additional constraint                     & 3.0\% & 11.9\% & 0.0040 & 1.0\% \\
new grid + $L$ as additional constraint & 4.0\% & 11.9\% & 0.0053 & 1.4\% \\
new grid + two additional constraints                       & 2.9\%                         & 11.4\%                         & 0.0039                         & 1.0\%                         \\ \hline
\end{tabular}
\end{table}

\subsection{Modelling-based Scaling Relations}

LI22 used modeling-inferred masses and radii to correct the scaling relations. With improved stellar parameters, we have carried out a similar analysis. The first scaling relation, in its standard form, is that \numax{} is proportional to $g T_{\mathrm{eff}}^{-0.5}$ \citep{1991ApJ...368..599B, 1995A&A...293...87K}. We fitted the \numax{} scaling relation using observed \numax{}, \teff{}, {\rm [M/H]}, and modeling-inferred \logg{}. 
For \apogee{} and \lamost{} targets, we derived following results
\begin{equation}
    \frac{\nu_{\rm max }}{\nu_{\rm max , \odot}}=\frac{g_{\mathrm{fit}}}{g_{\odot}}\left(\frac{T_{\mathrm{eff}}}{T_{\mathrm{eff}, \odot}}\right)^{-0.397}\left(10^{[\mathrm{M} / \mathrm{H}]}\right)^{-0.008} \text { (APOGEE) },
\end{equation}
and 
\begin{equation}
    \frac{\nu_{\rm max }}{\nu_{\rm max , \odot}}=\frac{g_{\mathrm{fit}}}{g_{\odot}}\left(\frac{T_{\mathrm{eff}}}{T_{\mathrm{eff}, \odot}}\right)^{-0.343}\left(10^{[\mathrm{M} / \mathrm{H}]}\right)^{-0.012} \text { (LAMOST) },
\end{equation}
where the solar values are $\nu_{\rm max, \odot}$ = 3090 $\mu$Hz, $T_{\rm eff, \odot}$ = 5777 K, and $\log g_{\odot}$ = 4.44 \citep{2011ApJ...743..143H}.
Compared to the scaling relations in LI22, we found that the same power law for the $g$ term, but the exponents for the $T_{eff}$ and {\rm [M/H]} terms are marginally different (LI22 gave $-0.459$ and $-0.022$ for APOGEE; $-0.421$ and $-0.039$ for LAMOST). Our updated version suggests values closer to the standard scaling relation, with very little dependence on metallicity.

The second standard scaling relation is that the large frequency separation of radial modes, \Dnu{}, is proportional to the square root of the mean stellar density \citep{1986ApJ...306L..37U}. We used the observed \Dnu{} and the model-determined mean density to fit the \Dnu{} scaling relation and obtained
\begin{equation}
    \frac{\Delta \nu}{\Delta \nu_{\odot}}=\left(\frac{\bar{\rho}_{\mathrm{fit}}}{\bar{\rho}_{\odot}}\right)^{0.507},
\end{equation}
where $\Delta \nu_{\rm \odot}$ = 135.1 $\mu$Hz.
This result is identical to that given by LI22, indicating that the \Dnu{} scaling relation is not sensitive to the systematic offset in estimated masses. 

\begin{figure}[b]
     \centering
     \includegraphics[width=8.8cm]{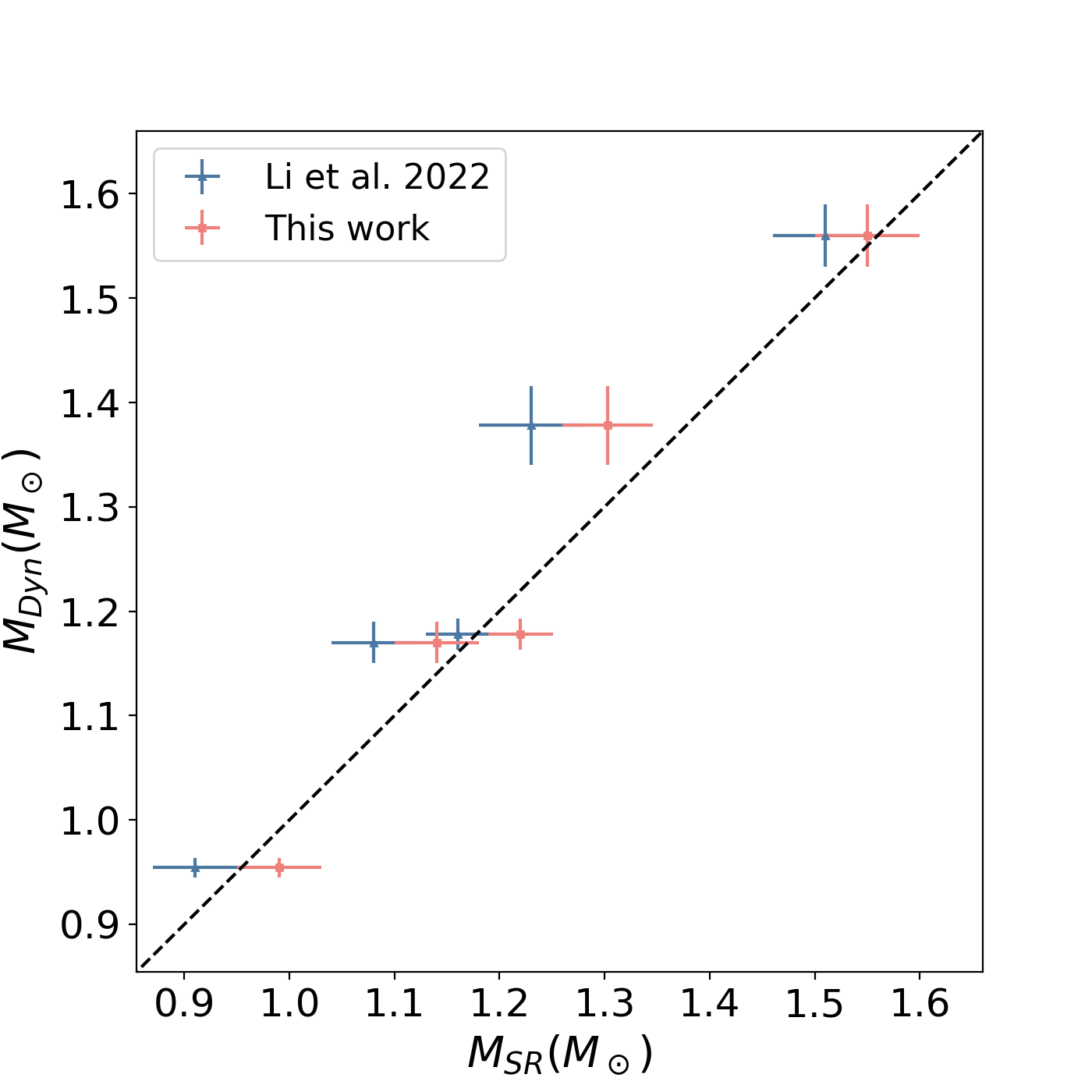}
     \includegraphics[width=8.8cm]{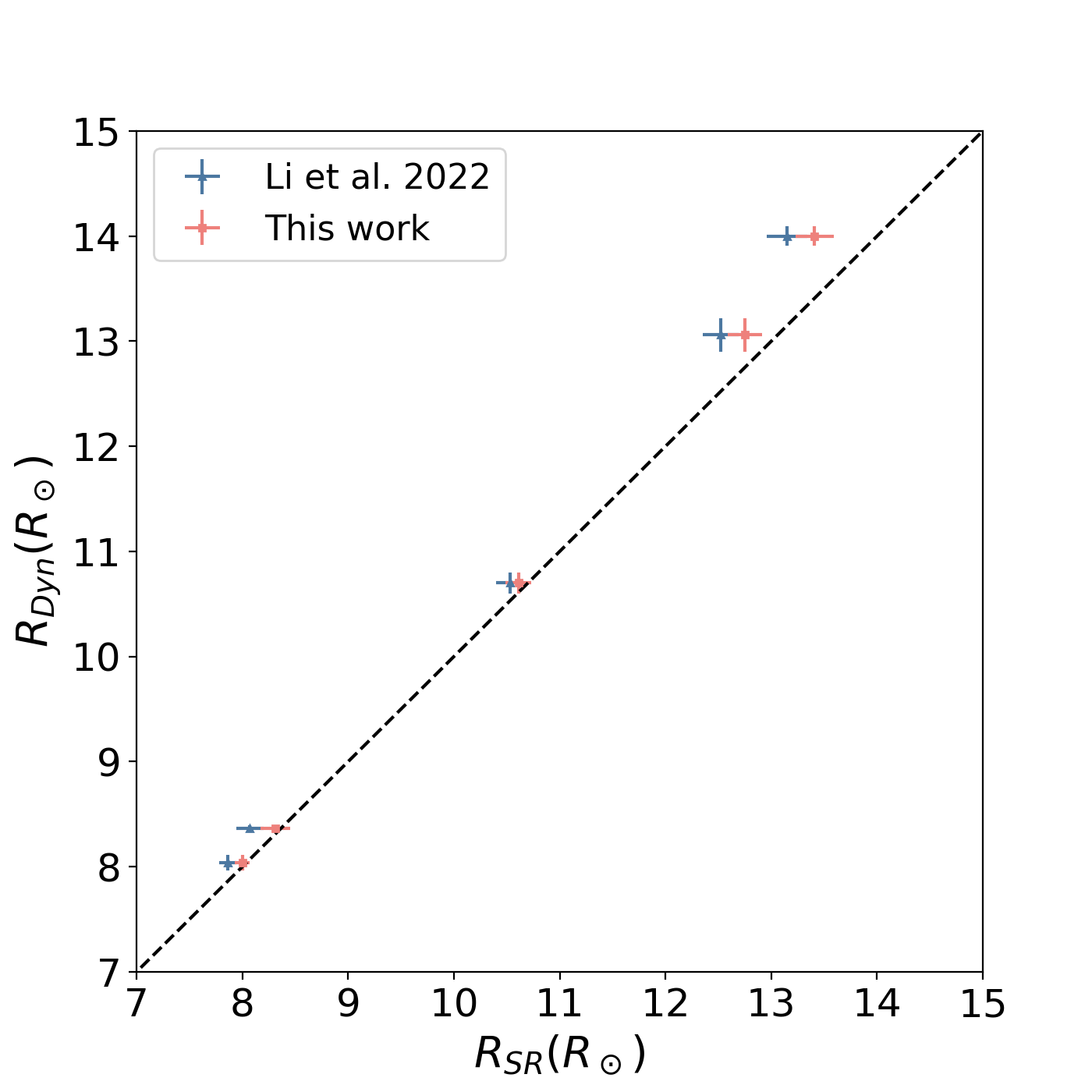}
     \caption{Comparison of the masses and radii of five red giants in binary systems derived from dynamical models and corrected scaling relations. The left panel shows the difference between the two methods for the masses of each star. The right panel shows the same for the radii.  The blue points use the corrected scaling relations from LI22, while the red points use the corrected scaling relations from this work.}
     \label{fig: 6}
\end{figure}

\begin{figure}
     \centering
     \includegraphics[width=8.8cm]{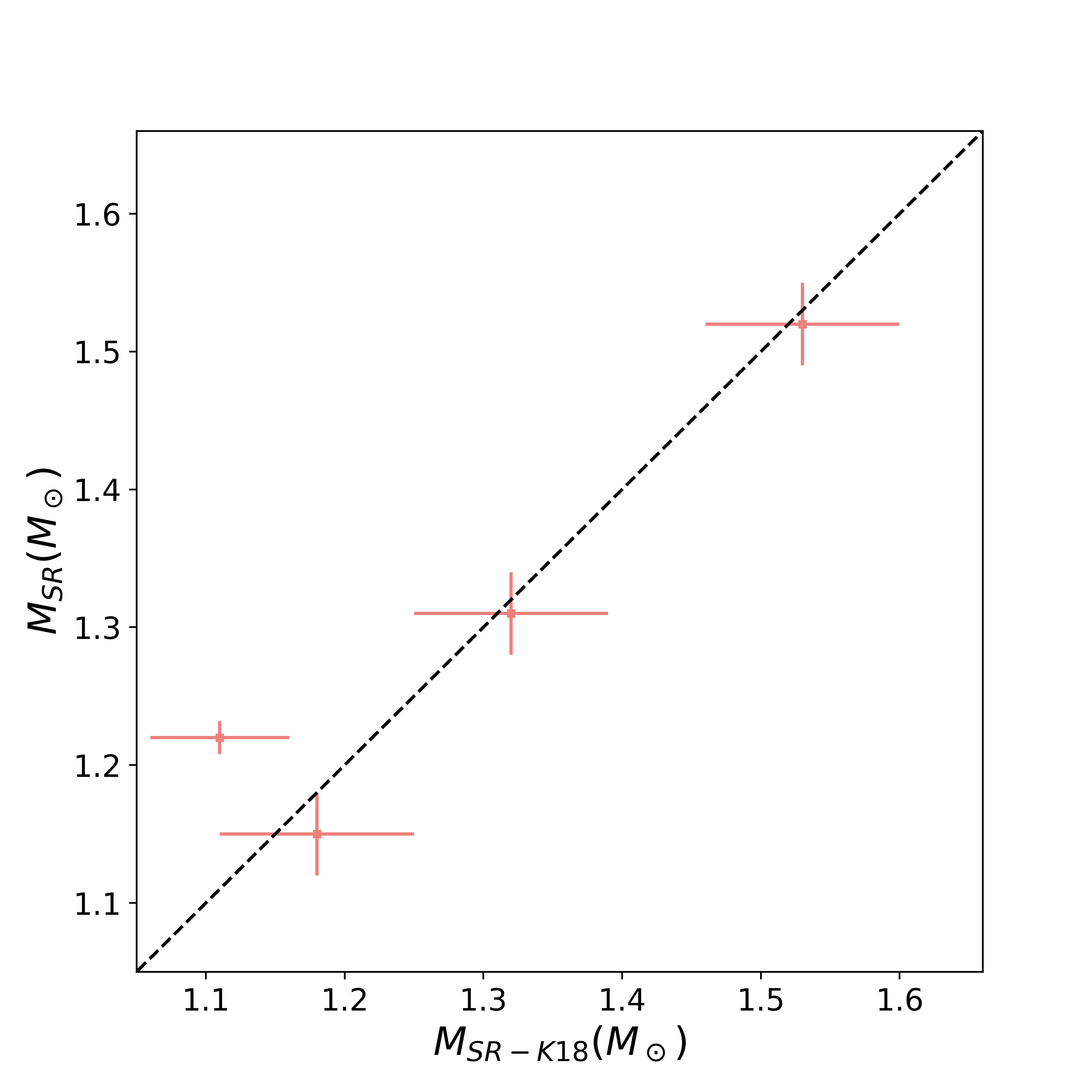}
     \includegraphics[width=8.8cm]{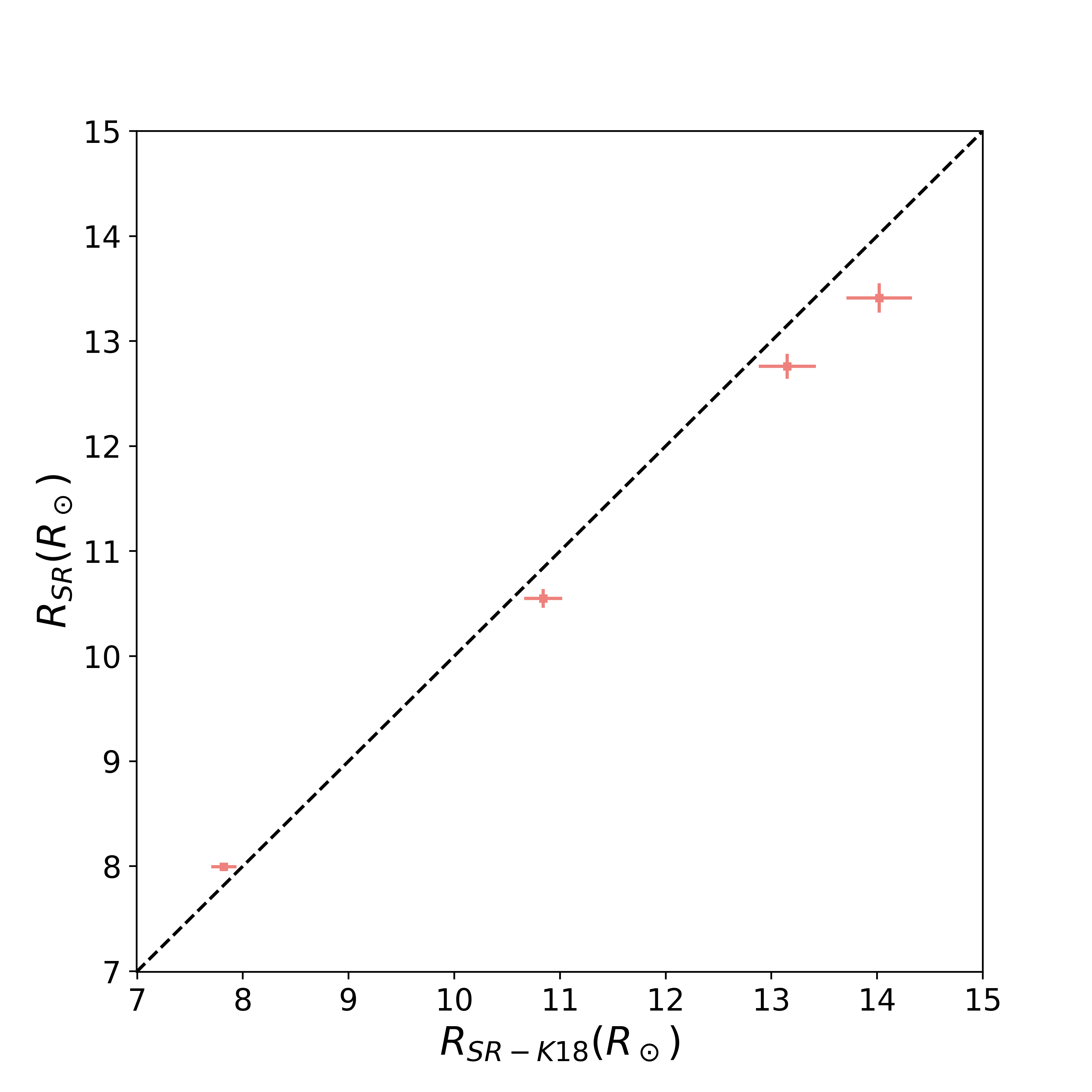}
     \caption{Comparison of the masses and radii of four red giants in binary systems derived from and corrected scaling relations. The left panel shows the difference between the two methods for the masses of each star. The right panel shows the same for the radii.  The horizontal coordinate uses the corrected scaling relations from \cite{2018A&A...616A.104K}, while the ordinate uses the corrected scaling relations from this work.}
     \label{fig: 7}
\end{figure}

To test the accuracy of the scaling relations after applying corrections for various effects, we used a sample of five red giants in eclipsing binary (EB) systems, whose masses and radii are accurately measured by dynamical modeling. The five red giants are KIC~8410637 \citep{frandsen2013kic}; KIC~9970396, KIC~7037405, KIC~9540226 \citep{brogaard2018establishing}, and KIC~4054905 \citep{brogaard2022establishing}. They all have high-resolution spectra from APOGEE, allowing us to use the corrected scaling relations calibrated for APOGEE data. We compared our results with those from the corrected scaling relations of LI22. Figure~\ref{fig: 6} shows that our new scaling relations yield masses within 1$\sigma$ of the dynamical masses for three red giants and within 1.5$\sigma$ for the other two, and significantly improve the accuracy of mass and radius estimation compared to LI22's scaling relations.
\cite{2018A&A...616A.104K} used four of the mentioned eclipsing binary red giants, KIC~8410637; KIC~9970396, KIC~7037405, KIC~9540226. They used these stars to revise the scaling relations, so we use their results as a reference for our corrected scaling relations. Figure~\ref{fig: 7} shows that our new scaling relations yield masses within 1.5$\sigma$ of the seismic masses, using the corrected scaling relations from \cite{2018A&A...616A.104K} for these red giants.

\subsection{Surface Term}

\begin{figure}[t]
    \centering
    \includegraphics[width=\textwidth]{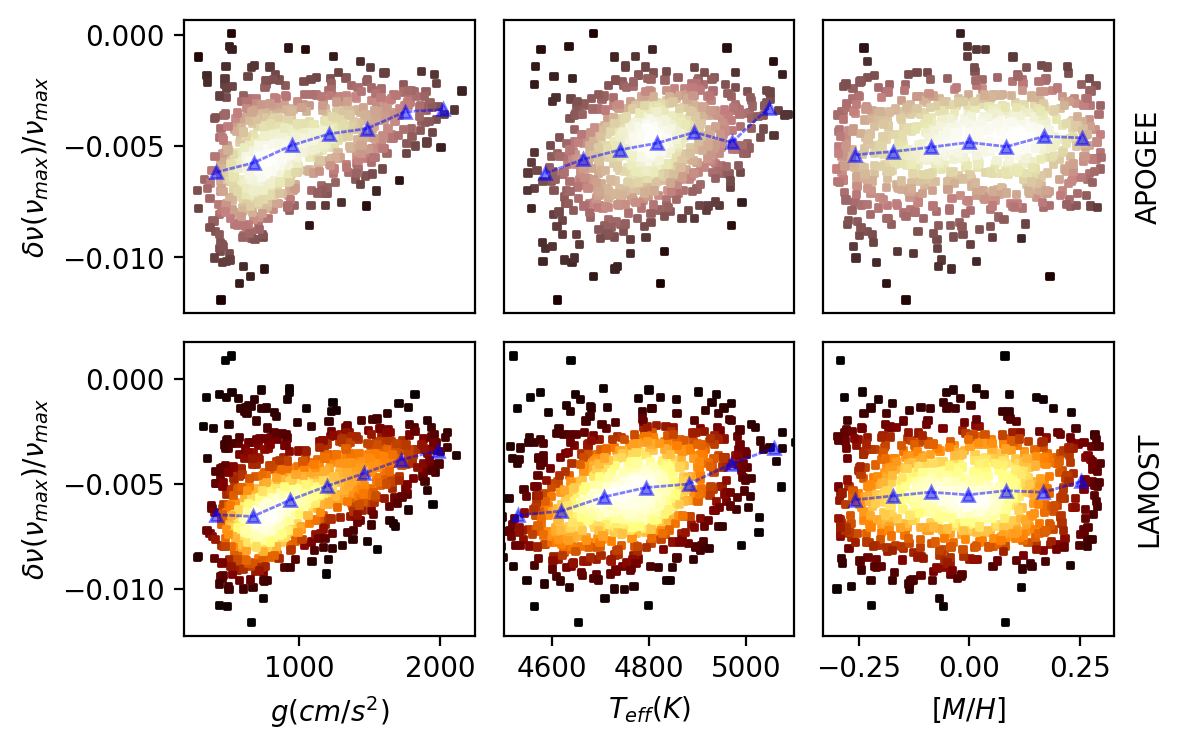}
    \caption{Density scatter plots of the surface correction with three stellar parameters. The top row of subplots shows stars observed by the \apogee{}, and the bottom row shows the \lamost{} targets. The filled triangle represents the middle value of each bin.}
    \label{fig: corner-apogee}
\end{figure}

The surface term in asteroseismology refers to the differences between observed oscillation frequencies and those of the best-fitting model \citep{10.1093/mnras/199.3.735}. The surface term is caused by incorrect modeling of the near-surface layers in stellar code. Given that the properties near-surface layers largely correlate to global parameters, the surface term is expected to vary smoothly as a function of effective temperature, surface gravity, and metallicity (\citealt{2017MNRAS.466L..43T}; \citealt{2018MNRAS.479.4416C}; \citealt{2020MNRAS.495.4965J}; \citealt{2021ApJ...906...54O}; \mycitealt{li2022prescription}). The star sample in this work makes it possible to systematically study the surface term and its dependencies on surface features in a wide parameter range.

We investigated the correlations between $\delta\nu(\numax)/\numax$ and three parameters, i.e., seismic surface gravity ($g$), effective temperature (\teff), and metallicity ([M/H]), in Figure~\ref{fig: corner-apogee}. We found that the surface term strongly depends on the surface gravity and effective temperature, but there is no significant correlation with metallicity.
We fitted $\delta\nu(\numax)/\numax$ as a function of two surface parameters using the formula as follows:
\begin{equation}
    \frac{\delta\nu(\numax)}{\numax} = \beta_0 \left(\frac{g}{g_{\odot}}\right)^{\beta_1} \left(\frac{T_{\mathrm{eff}}}{T_{\mathrm{eff},\odot}}\right)^{\beta_2}
\end{equation}
We used the {\tt scipy curve\_fit} module and found the best-fitting parameters are $[\beta_{0}, \beta_{1}, \beta_2] = [-0.0014 \pm 0.0002, -0.32 \pm 0.04, -1.1 \pm 0.7]$ with the \apogee{} \teff{} and [M/H], and $[\beta_{0}, \beta_{1}, \beta_2] = [-0.0016 \pm 0.0002, -0.26 \pm 0.03, -1.7 \pm 0.4]$ with the \lamost{} data. We also calculated the residual between fitting and true value. After analyzing the residual, we still failed to find any correlation with metallicity.
Moreover, the absolute value of $\delta\nu(\numax)/\numax$ is relatively large for stars with lower surface gravity, indicating that the surface effect increases with stars' evolution on the RGB. 

\section{Conclusions} \label{sec:4}

As a follow-up study of LI22, we have introduced two additional observed constraints to improve the estimated fundamental parameters of a sample of  \kepler{} red giants.
We notice that the gravity-mode period spacing and \gaia{} luminosity significantly improve the precision of 1,153 red giant branch stars. The typical uncertainty is 2.9$\%$ for the mass, 11$\%$ for the age, 1.0$\%$ for the radius, 0.0039 dex for the surface gravity, and 0.5$\%$ for the helium core mass, making this the best-characterized sample of red-giant stars available to date.

With the improved stellar parameters, we re-derive the seismic scaling relations. Compare with our previous version, the updated \numax{} scaling relation suggests a relatively small dependence on the effective temperature and the metallicity. 
Moreover, we systematically study the surface term for red giant stars. The results indicate that the surface term increases when stars become more evolved on RGB. The surface term strongly depends on the surface gravity and effective temperature, but we find no significant correlation with metallicity.  

\begin{acknowledgments}

This work is supported by the National Natural Science Foundation of China (NSFC) (grants 12090040, 12090042) and the Joint Research Fund in Astronomy (U2031203) under cooperative agreement between the National Natural Science Foundation of China (NSFC) and Chinese Academy of Sciences (CAS).
This work is also supported by the Fundamental Research Funds for the Central Universities.
This paper has also received funding from the European Research Council (ERC) under the European Union's Horizon 2020 research and innovation programme (CartographY GA. 804752). 
TRB acknowledges support from the Australian Research Council through Laureate Fellowship FL220100117.
We also thank the \kepler{} team for making this research possible.
\end{acknowledgments}

\bibliography{references/myastronomybib}

\begin{thebibliography}{}
\expandafter\ifx\csname natexlab\endcsname\relax\def\natexlab#1{#1}\fi
\providecommand{\url}[1]{\href{#1}{#1}}
\providecommand{\dodoi}[1]{doi:~\href{http://doi.org/#1}{\nolinkurl{#1}}}
\providecommand{\doeprint}[1]{\href{http://ascl.net/#1}{\nolinkurl{http://ascl.net/#1}}}
\providecommand{\doarXiv}[1]{\href{https://arxiv.org/abs/#1}{\nolinkurl{https://arxiv.org/abs/#1}}}

\bibitem[{{Ball} \& {Gizon}(2014)}]{2014A&A...568A.123B}
{Ball}, W.~H., \& {Gizon}, L. 2014, \aap, 568, A123,
  \dodoi{10.1051/0004-6361/201424325}

\bibitem[{Basu {et~al.}(2011)Basu, Grundahl, Stello, Kallinger, Hekker, Mosser,
  Garc{\'\i}a, Mathur, Brogaard, Bruntt, {et~al.}}]{basu2011sounding}
Basu, S., Grundahl, F., Stello, D., {et~al.} 2011, The Astrophysical Journal
  Letters, 729, L10, \dodoi{10.1088/2041-8205/729/1/L10}

\bibitem[{Bedding {et~al.}(2011)Bedding, Mosser, Huber, Montalbán, Beck,
  Christensen-Dalsgaard, Elsworth, García, Miglio, Stello, White, De~Ridder,
  Hekker, Aerts, Barban, Belkacem, Broomhall, Brown, Buzasi, Carrier, Chaplin,
  Di~Mauro, Dupret, Frandsen, Gilliland, Goupil, Jenkins, Kallinger, Kawaler,
  Kjeldsen, Mathur, Noels, Aguirre, \& Ventura}]{Bedding_2011}
Bedding, T.~R., Mosser, B., Huber, D., {et~al.} 2011, Nature, 471, 608,
  \dodoi{10.1038/nature09935}

\bibitem[{{Berger} {et~al.}(2020){Berger}, {Huber}, {van Saders}, {Gaidos},
  {Tayar}, \& {Kraus}}]{2020AJ....159..280B}
{Berger}, T.~A., {Huber}, D., {van Saders}, J.~L., {et~al.} 2020, \aj, 159,
  280, \dodoi{10.3847/1538-3881/159/6/280}

\bibitem[{Borucki {et~al.}(2008)Borucki, Koch, Batalha, Caldwell,
  Christensen-Dalsgaard, Cochran, Dunham, Gautier, Geary, Gilliland,
  {et~al.}}]{borucki2008kepler}
Borucki, W., Koch, D., Batalha, N., {et~al.} 2008, Proceedings of the
  International Astronomical Union, 4, 289.
\newblock \url{https://ui.adsabs.harvard.edu/abs/2009IAUS..253..289B}

\bibitem[{Brogaard {et~al.}(2018)Brogaard, Hansen, Miglio, Slumstrup, Frandsen,
  Jessen-Hansen, Lund, Bossini, Thygesen, Davies,
  {et~al.}}]{brogaard2018establishing}
Brogaard, K., Hansen, C., Miglio, A., {et~al.} 2018, Monthly Notices of the
  Royal Astronomical Society, 476, 3729, \dodoi{10.1093/mnras/sty268}

\bibitem[{Brogaard {et~al.}(2022)Brogaard, Arentoft, Slumstrup, Grundahl, Lund,
  Arndt, Grund, Rudrasingam, Theil, Christensen,
  {et~al.}}]{brogaard2022establishing}
Brogaard, K., Arentoft, T., Slumstrup, D., {et~al.} 2022, arXiv preprint
  arXiv:2210.02059, \dodoi{10.1051/0004-6361/202244345}

\bibitem[{{Brown} {et~al.}(1991){Brown}, {Gilliland}, {Noyes}, \&
  {Ramsey}}]{1991ApJ...368..599B}
{Brown}, T.~M., {Gilliland}, R.~L., {Noyes}, R.~W., \& {Ramsey}, L.~W. 1991,
  \apj, 368, 599, \dodoi{10.1086/169725}

\bibitem[{Chaplin \& Miglio(2013)}]{chaplin2013asteroseismology}
Chaplin, W.~J., \& Miglio, A. 2013, Annual Review of Astronomy and
  Astrophysics, 51, 353, \dodoi{10.1146/annurev-astro-082812-140938}

\bibitem[{{Choi} {et~al.}(2016){Choi}, {Dotter}, {Conroy}, {Cantiello},
  {Paxton}, \& {Johnson}}]{2016ApJ...823..102C}
{Choi}, J., {Dotter}, A., {Conroy}, C., {et~al.} 2016, \apj, 823, 102,
  \dodoi{10.3847/0004-637X/823/2/102}

\bibitem[{Christensen-Dalsgaard(1982)}]{10.1093/mnras/199.3.735}
Christensen-Dalsgaard, J. 1982, Monthly Notices of the Royal Astronomical
  Society, 199, 735, \dodoi{10.1093/mnras/199.3.735}

\bibitem[{{Compton} {et~al.}(2018){Compton}, {Bedding}, {Ball}, {Stello},
  {Huber}, {White}, \& {Kjeldsen}}]{2018MNRAS.479.4416C}
{Compton}, D.~L., {Bedding}, T.~R., {Ball}, W.~H., {et~al.} 2018, \mnras, 479,
  4416, \dodoi{10.1093/mnras/sty1632}

\bibitem[{Deheuvels {et~al.}(2012)Deheuvels, García, Chaplin, Basu, Antia,
  Appourchaux, Benomar, Davies, Elsworth, Gizon, Goupil, Reese, Regulo, Schou,
  Stahn, Casagrande, Christensen-Dalsgaard, Fischer, Hekker, Kjeldsen, Mathur,
  Mosser, Pinsonneault, Valenti, Christiansen, Kinemuchi, \&
  Mullally}]{Deheuvels_2012}
Deheuvels, S., García, R.~A., Chaplin, W.~J., {et~al.} 2012, The Astrophysical
  Journal, 756, 19, \dodoi{10.1088/0004-637X/756/1/19}

\bibitem[{{Deheuvels, S.} {et~al.}(2022){Deheuvels, S.}, {Ballot, J.}, {Gehan,
  C.}, \& {Mosser, B.}}]{S.Deheuvels_2022}
{Deheuvels, S.}, {Ballot, J.}, {Gehan, C.}, \& {Mosser, B.} 2022, A\&A, 659,
  A106, \dodoi{10.1051/0004-6361/202142094}

\bibitem[{{El-Badry} {et~al.}(2021){El-Badry}, {Rix}, \&
  {Heintz}}]{2021MNRAS.506.2269E}
{El-Badry}, K., {Rix}, H.-W., \& {Heintz}, T.~M. 2021, \mnras, 506, 2269,
  \dodoi{10.1093/mnras/stab323}

\bibitem[{Frandsen {et~al.}(2013)Frandsen, Lehmann, Hekker, Southworth,
  Debosscher, Beck, Hartmann, Pigulski, Kopacki, Ko{\l}aczkowski,
  {et~al.}}]{frandsen2013kic}
Frandsen, S., Lehmann, H., Hekker, S., {et~al.} 2013, Astronomy \&
  Astrophysics, 556, A138, \dodoi{10.1051/0004-6361/201321817}

\bibitem[{Gai {et~al.}(2011)Gai, Basu, Chaplin, \& Elsworth}]{gai2011depth}
Gai, N., Basu, S., Chaplin, W.~J., \& Elsworth, Y. 2011, The Astrophysical
  Journal, 730, 63, \dodoi{10.1088/0004-637X/730/2/63}

\bibitem[{{Gaia Collaboration} {et~al.}(2016){Gaia Collaboration}, {Prusti},
  {de Bruijne}, {Brown}, {Vallenari}, {Babusiaux}, {Bailer-Jones}, {Bastian},
  {Biermann}, {Evans}, {Eyer}, {Jansen}, {Jordi}, {Klioner}, {Lammers},
  {Lindegren}, {Luri}, {Mignard}, {Milligan}, {Panem}, {Poinsignon},
  {Pourbaix}, {Randich}, {Sarri}, {Sartoretti}, {Siddiqui}, {Soubiran},
  {Valette}, {van Leeuwen}, {Walton}, {Aerts}, {Arenou}, {Cropper}, {Drimmel},
  {H{\o}g}, {Katz}, {Lattanzi}, {O'Mullane}, {Grebel}, {Holland}, {Huc},
  {Passot}, {Bramante}, {Cacciari}, {Casta{\~n}eda}, {Chaoul}, {Cheek}, {De
  Angeli}, {Fabricius}, {Guerra}, {Hern{\'a}ndez}, {Jean-Antoine-Piccolo},
  {Masana}, {Messineo}, {Mowlavi}, {Nienartowicz}, {Ord{\'o}{\~n}ez-Blanco},
  {Panuzzo}, {Portell}, {Richards}, {Riello}, {Seabroke}, {Tanga},
  {Th{\'e}venin}, {Torra}, {Els}, {Gracia-Abril}, {Comoretto},
  {Garcia-Reinaldos}, {Lock}, {Mercier}, {Altmann}, {Andrae}, {Astraatmadja},
  {Bellas-Velidis}, {Benson}, {Berthier}, {Blomme}, {Busso}, {Carry},
  {Cellino}, {Clementini}, {Cowell}, {Creevey}, {Cuypers}, {Davidson}, {De
  Ridder}, {de Torres}, {Delchambre}, {Dell'Oro}, {Ducourant}, {Fr{\'e}mat},
  {Garc{\'\i}a-Torres}, {Gosset}, {Halbwachs}, {Hambly}, {Harrison}, {Hauser},
  {Hestroffer}, {Hodgkin}, {Huckle}, {Hutton}, {Jasniewicz}, {Jordan},
  {Kontizas}, {Korn}, {Lanzafame}, {Manteiga}, {Moitinho}, {Muinonen},
  {Osinde}, {Pancino}, {Pauwels}, {Petit}, {Recio-Blanco}, {Robin}, {Sarro},
  {Siopis}, {Smith}, {Smith}, {Sozzetti}, {Thuillot}, {van Reeven}, {Viala},
  {Abbas}, {Abreu Aramburu}, {Accart}, {Aguado}, {Allan}, {Allasia},
  {Altavilla}, {{\'A}lvarez}, {Alves}, {Anderson}, {Andrei}, {Anglada Varela},
  {Antiche}, {Antoja}, {Ant{\'o}n}, {Arcay}, {Atzei}, {Ayache}, {Bach},
  {Baker}, {Balaguer-N{\'u}{\~n}ez}, {Barache}, {Barata}, {Barbier}, {Barblan},
  {Baroni}, {Barrado y Navascu{\'e}s}, {Barros}, {Barstow}, {Becciani},
  {Bellazzini}, {Bellei}, {Bello Garc{\'\i}a}, {Belokurov}, {Bendjoya},
  {Berihuete}, {Bianchi}, {Bienaym{\'e}}, {Billebaud}, {Blagorodnova},
  {Blanco-Cuaresma}, {Boch}, {Bombrun}, {Borrachero}, {Bouquillon}, {Bourda},
  {Bouy}, {Bragaglia}, {Breddels}, {Brouillet}, {Br{\"u}semeister},
  {Bucciarelli}, {Budnik}, {Burgess}, {Burgon}, {Burlacu}, {Busonero}, {Buzzi},
  {Caffau}, {Cambras}, {Campbell}, {Cancelliere}, {Cantat-Gaudin}, {Carlucci},
  {Carrasco}, {Castellani}, {Charlot}, {Charnas}, {Charvet}, {Chassat},
  {Chiavassa}, {Clotet}, {Cocozza}, {Collins}, {Collins}, {Costigan}, {Crifo},
  {Cross}, {Crosta}, {Crowley}, {Dafonte}, {Damerdji}, {Dapergolas}, {David},
  {David}, {De Cat}, {de Felice}, {de Laverny}, {De Luise}, {De March}, {de
  Martino}, {de Souza}, {Debosscher}, {del Pozo}, {Delbo}, {Delgado},
  {Delgado}, {di Marco}, {Di Matteo}, {Diakite}, {Distefano}, {Dolding}, {Dos
  Anjos}, {Drazinos}, {Dur{\'a}n}, {Dzigan}, {Ecale}, {Edvardsson}, {Enke},
  {Erdmann}, {Escolar}, {Espina}, {Evans}, {Eynard Bontemps}, {Fabre},
  {Fabrizio}, {Faigler}, {Falc{\~a}o}, {Farr{\`a}s Casas}, {Faye}, {Federici},
  {Fedorets}, {Fern{\'a}ndez-Hern{\'a}ndez}, {Fernique}, {Fienga}, {Figueras},
  {Filippi}, {Findeisen}, {Fonti}, {Fouesneau}, {Fraile}, {Fraser}, {Fuchs},
  {Furnell}, {Gai}, {Galleti}, {Galluccio}, {Garabato}, {Garc{\'\i}a-Sedano},
  {Gar{\'e}}, {Garofalo}, {Garralda}, {Gavras}, {Gerssen}, {Geyer}, {Gilmore},
  {Girona}, {Giuffrida}, {Gomes}, {Gonz{\'a}lez-Marcos},
  {Gonz{\'a}lez-N{\'u}{\~n}ez}, {Gonz{\'a}lez-Vidal}, {Granvik}, {Guerrier},
  {Guillout}, {Guiraud}, {G{\'u}rpide}, {Guti{\'e}rrez-S{\'a}nchez}, {Guy},
  {Haigron}, {Hatzidimitriou}, {Haywood}, {Heiter}, {Helmi}, {Hobbs},
  {Hofmann}, {Holl}, {Holland}, {Hunt}, {Hypki}, {Icardi}, {Irwin}, {Jevardat
  de Fombelle}, {Jofr{\'e}}, {Jonker}, {Jorissen}, {Julbe}, {Karampelas},
  {Kochoska}, {Kohley}, {Kolenberg}, {Kontizas}, {Koposov}, {Kordopatis},
  {Koubsky}, {Kowalczyk}, {Krone-Martins}, {Kudryashova}, {Kull}, {Bachchan},
  {Lacoste-Seris}, {Lanza}, {Lavigne}, {Le Poncin-Lafitte}, {Lebreton},
  {Lebzelter}, {Leccia}, {Leclerc}, {Lecoeur-Taibi}, {Lemaitre}, {Lenhardt},
  {Leroux}, {Liao}, {Licata}, {Lindstr{\o}m}, {Lister}, {Livanou}, {Lobel},
  {L{\"o}ffler}, {L{\'o}pez}, {Lopez-Lozano}, {Lorenz}, {Loureiro},
  {MacDonald}, {Magalh{\~a}es Fernandes}, {Managau}, {Mann}, {Mantelet},
  {Marchal}, {Marchant}, {Marconi}, {Marie}, {Marinoni}, {Marrese},
  {Marschalk{\'o}}, {Marshall}, {Mart{\'\i}n-Fleitas}, {Martino}, {Mary},
  {Matijevi{\v{c}}}, {Mazeh}, {McMillan}, {Messina}, {Mestre}, {Michalik},
  {Millar}, {Miranda}, {Molina}, {Molinaro}, {Molinaro}, {Moln{\'a}r},
  {Moniez}, {Montegriffo}, {Monteiro}, {Mor}, {Mora}, {Morbidelli}, {Morel},
  {Morgenthaler}, {Morley}, {Morris}, {Mulone}, {Muraveva}, {Musella},
  {Narbonne}, {Nelemans}, {Nicastro}, {Noval}, {Ord{\'e}novic},
  {Ordieres-Mer{\'e}}, {Osborne}, {Pagani}, {Pagano}, {Pailler}, {Palacin},
  {Palaversa}, {Parsons}, {Paulsen}, {Pecoraro}, {Pedrosa}, {Pentik{\"a}inen},
  {Pereira}, {Pichon}, {Piersimoni}, {Pineau}, {Plachy}, {Plum}, {Poujoulet},
  {Pr{\v{s}}a}, {Pulone}, {Ragaini}, {Rago}, {Rambaux}, {Ramos-Lerate},
  {Ranalli}, {Rauw}, {Read}, {Regibo}, {Renk}, {Reyl{\'e}}, {Ribeiro},
  {Rimoldini}, {Ripepi}, {Riva}, {Rixon}, {Roelens}, {Romero-G{\'o}mez},
  {Rowell}, {Royer}, {Rudolph}, {Ruiz-Dern}, {Sadowski}, {Sagrist{\`a}
  Sell{\'e}s}, {Sahlmann}, {Salgado}, {Salguero}, {Sarasso}, {Savietto},
  {Schnorhk}, {Schultheis}, {Sciacca}, {Segol}, {Segovia}, {Segransan},
  {Serpell}, {Shih}, {Smareglia}, {Smart}, {Smith}, {Solano}, {Solitro},
  {Sordo}, {Soria Nieto}, {Souchay}, {Spagna}, {Spoto}, {Stampa}, {Steele},
  {Steidelm{\"u}ller}, {Stephenson}, {Stoev}, {Suess}, {S{\"u}veges}, {Surdej},
  {Szabados}, {Szegedi-Elek}, {Tapiador}, {Taris}, {Tauran}, {Taylor},
  {Teixeira}, {Terrett}, {Tingley}, {Trager}, {Turon}, {Ulla}, {Utrilla},
  {Valentini}, {van Elteren}, {Van Hemelryck}, {van Leeuwen}, {Varadi},
  {Vecchiato}, {Veljanoski}, {Via}, {Vicente}, {Vogt}, {Voss}, {Votruba},
  {Voutsinas}, {Walmsley}, {Weiler}, {Weingrill}, {Werner}, {Wevers},
  {Whitehead}, {Wyrzykowski}, {Yoldas}, {{\v{Z}}erjal}, {Zucker}, {Zurbach},
  {Zwitter}, {Alecu}, {Allen}, {Allende Prieto}, {Amorim},
  {Anglada-Escud{\'e}}, {Arsenijevic}, {Azaz}, {Balm}, {Beck}, {Bernstein},
  {Bigot}, {Bijaoui}, {Blasco}, {Bonfigli}, {Bono}, {Boudreault}, {Bressan},
  {Brown}, {Brunet}, {Bunclark}, {Buonanno}, {Butkevich}, {Carret}, {Carrion},
  {Chemin}, {Ch{\'e}reau}, {Corcione}, {Darmigny}, {de Boer}, {de Teodoro}, {de
  Zeeuw}, {Delle Luche}, {Domingues}, {Dubath}, {Fodor}, {Fr{\'e}zouls},
  {Fries}, {Fustes}, {Fyfe}, {Gallardo}, {Gallegos}, {Gardiol}, {Gebran},
  {Gomboc}, {G{\'o}mez}, {Grux}, {Gueguen}, {Heyrovsky}, {Hoar}, {Iannicola},
  {Isasi Parache}, {Janotto}, {Joliet}, {Jonckheere}, {Keil}, {Kim},
  {Klagyivik}, {Klar}, {Knude}, {Kochukhov}, {Kolka}, {Kos}, {Kutka}, {Lainey},
  {LeBouquin}, {Liu}, {Loreggia}, {Makarov}, {Marseille}, {Martayan},
  {Martinez-Rubi}, {Massart}, {Meynadier}, {Mignot}, {Munari}, {Nguyen},
  {Nordlander}, {Ocvirk}, {O'Flaherty}, {Olias Sanz}, {Ortiz}, {Osorio},
  {Oszkiewicz}, {Ouzounis}, {Palmer}, {Park}, {Pasquato}, {Peltzer}, {Peralta},
  {P{\'e}turaud}, {Pieniluoma}, {Pigozzi}, {Poels}, {Prat}, {Prod'homme},
  {Raison}, {Rebordao}, {Risquez}, {Rocca-Volmerange}, {Rosen}, {Ruiz-Fuertes},
  {Russo}, {Sembay}, {Serraller Vizcaino}, {Short}, {Siebert}, {Silva},
  {Sinachopoulos}, {Slezak}, {Soffel}, {Sosnowska}, {Strai{\v{z}}ys}, {ter
  Linden}, {Terrell}, {Theil}, {Tiede}, {Troisi}, {Tsalmantza}, {Tur},
  {Vaccari}, {Vachier}, {Valles}, {Van Hamme}, {Veltz}, {Virtanen}, {Wallut},
  {Wichmann}, {Wilkinson}, {Ziaeepour}, \& {Zschocke}}]{gaia-2016-mission}
{Gaia Collaboration}, {Prusti}, T., {de Bruijne}, J.~H.~J., {et~al.} 2016,
  \aap, 595, A1, \dodoi{10.1051/0004-6361/201629272}

\bibitem[{{Gaia Collaboration} {et~al.}(2021){Gaia Collaboration}, {Brown},
  {Vallenari}, {Prusti}, {de Bruijne}, {Babusiaux}, {Biermann}, {Creevey},
  {Evans}, {Eyer}, {Hutton}, {Jansen}, {Jordi}, {Klioner}, {Lammers},
  {Lindegren}, {Luri}, {Mignard}, {Panem}, {Pourbaix}, {Randich}, {Sartoretti},
  {Soubiran}, {Walton}, {Arenou}, {Bailer-Jones}, {Bastian}, {Cropper},
  {Drimmel}, {Katz}, {Lattanzi}, {van Leeuwen}, {Bakker}, {Cacciari},
  {Casta{\~n}eda}, {De Angeli}, {Ducourant}, {Fabricius}, {Fouesneau},
  {Fr{\'e}mat}, {Guerra}, {Guerrier}, {Guiraud}, {Jean-Antoine Piccolo},
  {Masana}, {Messineo}, {Mowlavi}, {Nicolas}, {Nienartowicz}, {Pailler},
  {Panuzzo}, {Riclet}, {Roux}, {Seabroke}, {Sordo}, {Tanga}, {Th{\'e}venin},
  {Gracia-Abril}, {Portell}, {Teyssier}, {Altmann}, {Andrae}, {Bellas-Velidis},
  {Benson}, {Berthier}, {Blomme}, {Brugaletta}, {Burgess}, {Busso}, {Carry},
  {Cellino}, {Cheek}, {Clementini}, {Damerdji}, {Davidson}, {Delchambre},
  {Dell'Oro}, {Fern{\'a}ndez-Hern{\'a}ndez}, {Galluccio}, {Garc{\'\i}a-Lario},
  {Garcia-Reinaldos}, {Gonz{\'a}lez-N{\'u}{\~n}ez}, {Gosset}, {Haigron},
  {Halbwachs}, {Hambly}, {Harrison}, {Hatzidimitriou}, {Heiter},
  {Hern{\'a}ndez}, {Hestroffer}, {Hodgkin}, {Holl}, {Jan{\ss}en}, {Jevardat de
  Fombelle}, {Jordan}, {Krone-Martins}, {Lanzafame}, {L{\"o}ffler}, {Lorca},
  {Manteiga}, {Marchal}, {Marrese}, {Moitinho}, {Mora}, {Muinonen}, {Osborne},
  {Pancino}, {Pauwels}, {Petit}, {Recio-Blanco}, {Richards}, {Riello},
  {Rimoldini}, {Robin}, {Roegiers}, {Rybizki}, {Sarro}, {Siopis}, {Smith},
  {Sozzetti}, {Ulla}, {Utrilla}, {van Leeuwen}, {van Reeven}, {Abbas}, {Abreu
  Aramburu}, {Accart}, {Aerts}, {Aguado}, {Ajaj}, {Altavilla}, {{\'A}lvarez},
  {{\'A}lvarez Cid-Fuentes}, {Alves}, {Anderson}, {Anglada Varela}, {Antoja},
  {Audard}, {Baines}, {Baker}, {Balaguer-N{\'u}{\~n}ez}, {Balbinot}, {Balog},
  {Barache}, {Barbato}, {Barros}, {Barstow}, {Bartolom{\'e}}, {Bassilana},
  {Bauchet}, {Baudesson-Stella}, {Becciani}, {Bellazzini}, {Bernet}, {Bertone},
  {Bianchi}, {Blanco-Cuaresma}, {Boch}, {Bombrun}, {Bossini}, {Bouquillon},
  {Bragaglia}, {Bramante}, {Breedt}, {Bressan}, {Brouillet}, {Bucciarelli},
  {Burlacu}, {Busonero}, {Butkevich}, {Buzzi}, {Caffau}, {Cancelliere},
  {C{\'a}novas}, {Cantat-Gaudin}, {Carballo}, {Carlucci}, {Carnerero},
  {Carrasco}, {Casamiquela}, {Castellani}, {Castro-Ginard}, {Castro Sampol},
  {Chaoul}, {Charlot}, {Chemin}, {Chiavassa}, {Cioni}, {Comoretto}, {Cooper},
  {Cornez}, {Cowell}, {Crifo}, {Crosta}, {Crowley}, {Dafonte}, {Dapergolas},
  {David}, {David}, {de Laverny}, {De Luise}, {De March}, {De Ridder}, {de
  Souza}, {de Teodoro}, {de Torres}, {del Peloso}, {del Pozo}, {Delbo},
  {Delgado}, {Delgado}, {Delisle}, {Di Matteo}, {Diakite}, {Diener},
  {Distefano}, {Dolding}, {Eappachen}, {Edvardsson}, {Enke}, {Esquej}, {Fabre},
  {Fabrizio}, {Faigler}, {Fedorets}, {Fernique}, {Fienga}, {Figueras},
  {Fouron}, {Fragkoudi}, {Fraile}, {Franke}, {Gai}, {Garabato},
  {Garcia-Gutierrez}, {Garc{\'\i}a-Torres}, {Garofalo}, {Gavras}, {Gerlach},
  {Geyer}, {Giacobbe}, {Gilmore}, {Girona}, {Giuffrida}, {Gomel}, {Gomez},
  {Gonzalez-Santamaria}, {Gonz{\'a}lez-Vidal}, {Granvik},
  {Guti{\'e}rrez-S{\'a}nchez}, {Guy}, {Hauser}, {Haywood}, {Helmi}, {Hidalgo},
  {Hilger}, {H{\l}adczuk}, {Hobbs}, {Holland}, {Huckle}, {Jasniewicz},
  {Jonker}, {Juaristi Campillo}, {Julbe}, {Karbevska}, {Kervella}, {Khanna},
  {Kochoska}, {Kontizas}, {Kordopatis}, {Korn}, {Kostrzewa-Rutkowska},
  {Kruszy{\'n}ska}, {Lambert}, {Lanza}, {Lasne}, {Le Campion}, {Le Fustec},
  {Lebreton}, {Lebzelter}, {Leccia}, {Leclerc}, {Lecoeur-Taibi}, {Liao},
  {Licata}, {Lindstr{\o}m}, {Lister}, {Livanou}, {Lobel}, {Madrero Pardo},
  {Managau}, {Mann}, {Marchant}, {Marconi}, {Marcos Santos}, {Marinoni},
  {Marocco}, {Marshall}, {Martin Polo}, {Mart{\'\i}n-Fleitas}, {Masip},
  {Massari}, {Mastrobuono-Battisti}, {Mazeh}, {McMillan}, {Messina},
  {Michalik}, {Millar}, {Mints}, {Molina}, {Molinaro}, {Moln{\'a}r},
  {Montegriffo}, {Mor}, {Morbidelli}, {Morel}, {Morris}, {Mulone}, {Munoz},
  {Muraveva}, {Murphy}, {Musella}, {Noval}, {Ord{\'e}novic}, {Orr{\`u}},
  {Osinde}, {Pagani}, {Pagano}, {Palaversa}, {Palicio}, {Panahi}, {Pawlak},
  {Pe{\~n}alosa Esteller}, {Penttil{\"a}}, {Piersimoni}, {Pineau}, {Plachy},
  {Plum}, {Poggio}, {Poretti}, {Poujoulet}, {Pr{\v{s}}a}, {Pulone}, {Racero},
  {Ragaini}, {Rainer}, {Raiteri}, {Rambaux}, {Ramos}, {Ramos-Lerate}, {Re
  Fiorentin}, {Regibo}, {Reyl{\'e}}, {Ripepi}, {Riva}, {Rixon}, {Robichon},
  {Robin}, {Roelens}, {Rohrbasser}, {Romero-G{\'o}mez}, {Rowell}, {Royer},
  {Rybicki}, {Sadowski}, {Sagrist{\`a} Sell{\'e}s}, {Sahlmann}, {Salgado},
  {Salguero}, {Samaras}, {Sanchez Gimenez}, {Sanna}, {Santove{\~n}a},
  {Sarasso}, {Schultheis}, {Sciacca}, {Segol}, {Segovia}, {S{\'e}gransan},
  {Semeux}, {Shahaf}, {Siddiqui}, {Siebert}, {Siltala}, {Slezak}, {Smart},
  {Solano}, {Solitro}, {Souami}, {Souchay}, {Spagna}, {Spoto}, {Steele},
  {Steidelm{\"u}ller}, {Stephenson}, {S{\"u}veges}, {Szabados}, {Szegedi-Elek},
  {Taris}, {Tauran}, {Taylor}, {Teixeira}, {Thuillot}, {Tonello}, {Torra},
  {Torra}, {Turon}, {Unger}, {Vaillant}, {van Dillen}, {Vanel}, {Vecchiato},
  {Viala}, {Vicente}, {Voutsinas}, {Weiler}, {Wevers}, {Wyrzykowski}, {Yoldas},
  {Yvard}, {Zhao}, {Zorec}, {Zucker}, {Zurbach}, \& {Zwitter}}]{gaia-2020-edr3}
{Gaia Collaboration}, {Brown}, A.~G.~A., {Vallenari}, A., {et~al.} 2021, \aap,
  649, A1, \dodoi{10.1051/0004-6361/202039657}

\bibitem[{{Green} {et~al.}(2019){Green}, {Schlafly}, {Zucker}, {Speagle}, \&
  {Finkbeiner}}]{2019ApJ...887...93G}
{Green}, G.~M., {Schlafly}, E., {Zucker}, C., {Speagle}, J.~S., \&
  {Finkbeiner}, D. 2019, \apj, 887, 93, \dodoi{10.3847/1538-4357/ab5362}

\bibitem[{Hekker \& Christensen-Dalsgaard(2017)}]{hekker2017giant}
Hekker, S., \& Christensen-Dalsgaard, J. 2017, The Astronomy and Astrophysics
  Review, 25, 1, \dodoi{10.1007/s00159-017-0101-x}

\bibitem[{{Hill} {et~al.}(2021){Hill}, {Kane}, {Campante}, {Li}, {Dalba},
  {Brandt}, {White}, {Pope}, {Stassun}, {Fulton}, {Corsaro}, {Li}, {Ong},
  {Bedding}, {Bossini}, {Buzasi}, {Chaplin}, {Cunha}, {Garc{\'\i}a}, {Breton},
  {Hon}, {Huber}, {Jiang}, {Kayhan}, {Kuszlewicz}, {Mathur}, {Serenelli}, \&
  {Stello}}]{2021AJ....162..211H}
{Hill}, M.~L., {Kane}, S.~R., {Campante}, T.~L., {et~al.} 2021, \aj, 162, 211,
  \dodoi{10.3847/1538-3881/ac1b31}

\bibitem[{{Hon} {et~al.}(2018){Hon}, {Stello}, \& {Yu}}]{2018MNRAS.476.3233H}
{Hon}, M., {Stello}, D., \& {Yu}, J. 2018, \mnras, 476, 3233,
  \dodoi{10.1093/mnras/sty483}

\bibitem[{{Huber} {et~al.}(2011){Huber}, {Bedding}, {Stello}, {Hekker},
  {Mathur}, {Mosser}, {Verner}, {Bonanno}, {Buzasi}, {Campante}, {Elsworth},
  {Hale}, {Kallinger}, {Silva Aguirre}, {Chaplin}, {De Ridder}, {Garc{\'\i}a},
  {Appourchaux}, {Frandsen}, {Houdek}, {Molenda-{\.Z}akowicz}, {Monteiro},
  {Christensen-Dalsgaard}, {Gilliland}, {Kawaler}, {Kjeldsen}, {Broomhall},
  {Corsaro}, {Salabert}, {Sanderfer}, {Seader}, \&
  {Smith}}]{2011ApJ...743..143H}
{Huber}, D., {Bedding}, T.~R., {Stello}, D., {et~al.} 2011, \apj, 743, 143,
  \dodoi{10.1088/0004-637X/743/2/143}

\bibitem[{{Huber} {et~al.}(2017){Huber}, {Zinn}, {Bojsen-Hansen},
  {Pinsonneault}, {Sahlholdt}, {Serenelli}, {Silva Aguirre}, {Stassun},
  {Stello}, {Tayar}, {Bastien}, {Bedding}, {Buchhave}, {Chaplin}, {Davies},
  {Garc{\'\i}a}, {Latham}, {Mathur}, {Mosser}, \&
  {Sharma}}]{2017ApJ...844..102H}
{Huber}, D., {Zinn}, J., {Bojsen-Hansen}, M., {et~al.} 2017, \apj, 844, 102,
  \dodoi{10.3847/1538-4357/aa75ca}

\bibitem[{{Huber} {et~al.}(2019){Huber}, {Chaplin}, {Chontos}, {Kjeldsen},
  {Christensen-Dalsgaard}, {Bedding}, {Ball}, {Brahm}, {Espinoza}, {Henning},
  {Jord{\'a}n}, {Sarkis}, {Knudstrup}, {Albrecht}, {Grundahl}, {Fredslund
  Andersen}, {Pall{\'e}}, {Crossfield}, {Fulton}, {Howard}, {Isaacson},
  {Weiss}, {Handberg}, {Lund}, {Serenelli}, {R{\o}rsted Mosumgaard},
  {Stokholm}, {Bieryla}, {Buchhave}, {Latham}, {Quinn}, {Gaidos}, {Hirano},
  {Ricker}, {Vanderspek}, {Seager}, {Jenkins}, {Winn}, {Antia}, {Appourchaux},
  {Basu}, {Bell}, {Benomar}, {Bonanno}, {Buzasi}, {Campante}, {{\c{C}}elik
  Orhan}, {Corsaro}, {Cunha}, {Davies}, {Deheuvels}, {Grunblatt}, {Hasanzadeh},
  {Di Mauro}, {Garc{\'\i}a}, {Gaulme}, {Girardi}, {Guzik}, {Hon}, {Jiang},
  {Kallinger}, {Kawaler}, {Kuszlewicz}, {Lebreton}, {Li}, {Lucas}, {Lundkvist},
  {Mann}, {Mathis}, {Mathur}, {Mazumdar}, {Metcalfe}, {Miglio}, {Monteiro},
  {Mosser}, {Noll}, {Nsamba}, {Ong}, {{\"O}rtel}, {Pereira}, {Ranadive},
  {R{\'e}gulo}, {Rodrigues}, {Roxburgh}, {Silva Aguirre}, {Smalley},
  {Schofield}, {Sousa}, {Stassun}, {Stello}, {Tayar}, {White}, {Verma},
  {Vrard}, {Y{\i}ld{\i}z}, {Baker}, {Bazot}, {Beichmann}, {Bergmann}, {Bugnet},
  {Cale}, {Carlino}, {Cartwright}, {Christiansen}, {Ciardi}, {Creevey},
  {Dittmann}, {Do Nascimento}, {Van Eylen}, {F{\"u}r{\'e}sz}, {Gagn{\'e}},
  {Gao}, {Gazeas}, {Giddens}, {Hall}, {Hekker}, {Ireland}, {Latouf}, {LeBrun},
  {Levine}, {Matzko}, {Natinsky}, {Page}, {Plavchan}, {Mansouri-Samani},
  {McCauliff}, {Mullally}, {Orenstein}, {Garcia Soto}, {Paegert}, {van Saders},
  {Schnaible}, {Soderblom}, {Szab{\'o}}, {Tanner}, {Tinney}, {Teske}, {Thomas},
  {Trampedach}, {Wright}, {Yuan}, \& {Zohrabi}}]{2019AJ....157..245H}
{Huber}, D., {Chaplin}, W.~J., {Chontos}, A., {et~al.} 2019, \aj, 157, 245,
  \dodoi{10.3847/1538-3881/ab1488}

\bibitem[{Jackiewicz(2021)}]{jackiewicz2021solar}
Jackiewicz, J. 2021, Frontiers in Astronomy and Space Sciences, 7, 595017,
  \dodoi{10.3389/fspas.2020.595017}

\bibitem[{{J{\o}rgensen} {et~al.}(2020){J{\o}rgensen}, {Montalb{\'a}n},
  {Miglio}, {Rendle}, {Davies}, {Buldgen}, {Scuflaire}, {Noels}, {Gaulme}, \&
  {Garc{\'\i}a}}]{2020MNRAS.495.4965J}
{J{\o}rgensen}, A. C.~S., {Montalb{\'a}n}, J., {Miglio}, A., {et~al.} 2020,
  \mnras, 495, 4965, \dodoi{10.1093/mnras/staa1480}

\bibitem[{{Kallinger} {et~al.}(2018){Kallinger}, {Beck}, {Stello}, \&
  {Garcia}}]{2018A&A...616A.104K}
{Kallinger}, T., {Beck}, P.~G., {Stello}, D., \& {Garcia}, R.~A. 2018, \aap,
  616, A104, \dodoi{10.1051/0004-6361/201832831}

\bibitem[{Kallinger {et~al.}(2010)Kallinger, Mosser, Hekker, Huber, Stello,
  Mathur, Basu, Bedding, Chaplin, De~Ridder,
  {et~al.}}]{kallinger2010asteroseismology}
Kallinger, T., Mosser, B., Hekker, S., {et~al.} 2010, Astronomy \&
  Astrophysics, 522, A1, \dodoi{10.1051/0004-6361/201015263}

\bibitem[{{Kallinger, T.} {et~al.}(2008){Kallinger, T.}, {Guenther, D. B.},
  {Matthews, J. M.}, {Weiss, W. W.}, {Huber, D.}, {Kuschnig, R.}, {Moffat, A.
  F. J.}, {Rucinski, S. M.}, \& {Sasselov, D.}}]{Kallinger_2008}
{Kallinger, T.}, {Guenther, D. B.}, {Matthews, J. M.}, {et~al.} 2008, A\&A,
  478, 497, \dodoi{10.1051/0004-6361:20078171}

\bibitem[{{Kjeldsen} \& {Bedding}(1995)}]{1995A&A...293...87K}
{Kjeldsen}, H., \& {Bedding}, T.~R. 1995, \aap, 293, 87,
  \dodoi{10.48550/arXiv.astro-ph/9403015}

\bibitem[{Li {et~al.}(2017)Li, Bedding, Huber, Ball, Stello, Murphy, \&
  Bland-Hawthorn}]{Li_2017}
Li, T., Bedding, T.~R., Huber, D., {et~al.} 2017, Monthly Notices of the Royal
  Astronomical Society, 475, 981, \dodoi{10.1093/mnras/stx3079}

\bibitem[{Li {et~al.}(2022{\natexlab{a}})Li, Li, Bi, Bedding, Davies, \&
  Du}]{Li_2022}
Li, T., Li, Y., Bi, S., {et~al.} 2022{\natexlab{a}}, The Astrophysical Journal,
  927, 167, \dodoi{10.3847/1538-4357/ac4fbf}

\bibitem[{Li {et~al.}(2022{\natexlab{b}})Li, Bedding, Stello, Huber, Hon,
  Joyce, Li, Perkins, White, Zinn, {et~al.}}]{li2022prescription}
Li, Y., Bedding, T.~R., Stello, D., {et~al.} 2022{\natexlab{b}}, arXiv preprint
  arXiv:2208.01176

\bibitem[{{Lindegren} {et~al.}(2021){Lindegren}, {Bastian}, {Biermann},
  {Bombrun}, {de Torres}, {Gerlach}, {Geyer}, {Hern{\'a}ndez}, {Hilger},
  {Hobbs}, {Klioner}, {Lammers}, {McMillan}, {Ramos-Lerate},
  {Steidelm{\"u}ller}, {Stephenson}, \& {van Leeuwen}}]{2021A&A...649A...4L}
{Lindegren}, L., {Bastian}, U., {Biermann}, M., {et~al.} 2021, \aap, 649, A4,
  \dodoi{10.1051/0004-6361/202039653}

\bibitem[{{Ma{\'\i}z Apell{\'a}niz} {et~al.}(2021){Ma{\'\i}z Apell{\'a}niz},
  {Pantaleoni Gonz{\'a}lez}, \& {Barb{\'a}}}]{2021A&A...649A..13M}
{Ma{\'\i}z Apell{\'a}niz}, J., {Pantaleoni Gonz{\'a}lez}, M., \& {Barb{\'a}},
  R.~H. 2021, \aap, 649, A13, \dodoi{10.1051/0004-6361/202140418}

\bibitem[{Metcalfe {et~al.}(2010)Metcalfe, Monteiro, Thompson,
  Molenda-{\.Z}akowicz, Appourchaux, Chaplin, Do{\u{g}}an, Eggenberger,
  Bedding, Bruntt, {et~al.}}]{metcalfe2010precise}
Metcalfe, T.~S., Monteiro, M., Thompson, M.~J., {et~al.} 2010, The
  Astrophysical Journal, 723, 1583, \dodoi{10.1088/0004-637X/723/2/1583}

\bibitem[{{Montalb{\'a}n} {et~al.}(2013){Montalb{\'a}n}, {Miglio}, {Noels},
  {Dupret}, {Scuflaire}, \& {Ventura}}]{2013ApJ...766..118M}
{Montalb{\'a}n}, J., {Miglio}, A., {Noels}, A., {et~al.} 2013, \apj, 766, 118,
  \dodoi{10.1088/0004-637X/766/2/118}

\bibitem[{{Mosser, B.} {et~al.}(2011){Mosser, B.}, {Barban, C.}, {Montalbán,
  J.}, {Beck, P. G.}, {Miglio, A.}, {Belkacem, K.}, {Goupil, M. J.}, {Hekker,
  S.}, {De Ridder, J.}, {Dupret, M. A.}, {Elsworth, Y.}, {Noels, A.}, {Baudin,
  F.}, {Michel, E.}, {Samadi, R.}, {Auvergne, M.}, {Baglin, A.}, \& {Catala,
  C.}}]{Mosser_2011}
{Mosser, B.}, {Barban, C.}, {Montalbán, J.}, {et~al.} 2011, A\&A, 532, A86,
  \dodoi{10.1051/0004-6361/201116825}

\bibitem[{{Murphy} {et~al.}(2021){Murphy}, {Li}, {Sekaran}, {Bedding}, {Yu},
  {Tkachenko}, {Colman}, {Huber}, {Hey}, {Baratashvili}, \&
  {Janssens}}]{2021MNRAS.505.2336M}
{Murphy}, S.~J., {Li}, T., {Sekaran}, S., {et~al.} 2021, \mnras, 505, 2336,
  \dodoi{10.1093/mnras/stab1436}

\bibitem[{{Ong} {et~al.}(2021){Ong}, {Basu}, \&
  {McKeever}}]{2021ApJ...906...54O}
{Ong}, J.~M.~J., {Basu}, S., \& {McKeever}, J.~M. 2021, \apj, 906, 54,
  \dodoi{10.3847/1538-4357/abc7c1}

\bibitem[{{P{\'e}rez Hern{\'a}ndez} {et~al.}(2016){P{\'e}rez Hern{\'a}ndez},
  {Garc{\'\i}a}, {Corsaro}, {Triana}, \& {De Ridder}}]{2016A&A...591A..99P}
{P{\'e}rez Hern{\'a}ndez}, F., {Garc{\'\i}a}, R.~A., {Corsaro}, E., {Triana},
  S.~A., \& {De Ridder}, J. 2016, \aap, 591, A99,
  \dodoi{10.1051/0004-6361/201628311}

\bibitem[{{Rui} \& {Fuller}(2021)}]{2021MNRAS.508.1618R}
{Rui}, N.~Z., \& {Fuller}, J. 2021, \mnras, 508, 1618,
  \dodoi{10.1093/mnras/stab2528}

\bibitem[{Silva~Aguirre {et~al.}(2015)Silva~Aguirre, Davies, Basu,
  Christensen-Dalsgaard, Creevey, Metcalfe, Bedding, Casagrande, Handberg,
  Lund, {et~al.}}]{silva2015ages}
Silva~Aguirre, V., Davies, G., Basu, S., {et~al.} 2015, Monthly Notices of the
  Royal Astronomical Society, 452, 2127, \dodoi{10.1093/mnras/stv1388}

\bibitem[{Stello {et~al.}(2009)Stello, Chaplin, Bruntt, Creevey,
  Garc{\'\i}a-Hern{\'a}ndez, Monteiro, Moya, Quirion, Sousa, Su{\'a}rez,
  {et~al.}}]{stello2009radius}
Stello, D., Chaplin, W.~J., Bruntt, H., {et~al.} 2009, The Astrophysical
  Journal, 700, 1589, \dodoi{10.1088/0004-637X/700/2/1589}

\bibitem[{Stello {et~al.}(2013)Stello, Huber, Bedding, Benomar, Bildsten,
  Elsworth, Gilliland, Mosser, Paxton, \& White}]{Stello_2013}
Stello, D., Huber, D., Bedding, T.~R., {et~al.} 2013, The Astrophysical Journal
  Letters, 765, L41, \dodoi{10.1088/2041-8205/765/2/L41}

\bibitem[{{Tayar} {et~al.}(2022){Tayar}, {Claytor}, {Huber}, \& {van
  Saders}}]{2022ApJ...927...31T}
{Tayar}, J., {Claytor}, Z.~R., {Huber}, D., \& {van Saders}, J. 2022, \apj,
  927, 31, \dodoi{10.3847/1538-4357/ac4bbc}

\bibitem[{{Trampedach} {et~al.}(2017){Trampedach}, {Aarslev}, {Houdek},
  {Collet}, {Christensen-Dalsgaard}, {Stein}, \&
  {Asplund}}]{2017MNRAS.466L..43T}
{Trampedach}, R., {Aarslev}, M.~J., {Houdek}, G., {et~al.} 2017, \mnras, 466,
  L43, \dodoi{10.1093/mnrasl/slw230}

\bibitem[{{Ulrich}(1986)}]{1986ApJ...306L..37U}
{Ulrich}, R.~K. 1986, \apjl, 306, L37, \dodoi{10.1086/184700}

\bibitem[{{Vrard, M.} {et~al.}(2016){Vrard, M.}, {Mosser, B.}, \& {Samadi,
  R.}}]{Vrard_2016}
{Vrard, M.}, {Mosser, B.}, \& {Samadi, R.} 2016, A\&A, 588, A87,
  \dodoi{10.1051/0004-6361/201527259}

\bibitem[{{Zhang} {et~al.}(2018){Zhang}, {Wu}, \& {Li}}]{2018ApJ...855...16Z}
{Zhang}, X., {Wu}, T., \& {Li}, Y. 2018, \apj, 855, 16,
  \dodoi{10.3847/1538-4357/aaaabb}

\bibitem[{{Zinn}(2021)}]{2021AJ....161..214Z}
{Zinn}, J.~C. 2021, \aj, 161, 214, \dodoi{10.3847/1538-3881/abe936}

\end{thebibliography}
\bibliographystyle{aasjournal} 

\end{document}